\documentclass[useAMS,usenatbib]{mn2e}

\usepackage{psfig}

\input psfig.sty
\usepackage{amsmath}
\usepackage{amssymb}
\usepackage{upgreek}
\usepackage{longtable}
\usepackage[mathscr]{eucal}

\usepackage{graphicx}

\newcommand{\arcs}{\mbox{\ensuremath{^{\prime\prime}}}}

\title[3D Photoionization Model of NGC 6302]
  {A 3D Photoionization Model of the Extreme Planetary Nebula NGC~6302}
\author[N.J.~Wright et al.]
  {N.J.~Wright,$^{1,2}$ M.J.~Barlow,$^2$ B.~Ercolano,$^{3}$ T.~Rauch$^4$\\
  $^1$Harvard-Smithsonian Center for Astrophysics, 60 Garden Street, Cambridge, MA 02138, USA\\
  $^2$Department of Physics and Astronomy, University College London, Gower Street, London, WC1E 6BT, UK\\
  $^3$Universit\"ats-Sternwarte M\"unchen, Scheinerstr. 1, 81679 M\"unchen, Germany\\
  $^4$Institute for Astronomy \& Astrophysics, Kepler Center for Astro \& Particle Physics, Eberhard Karls University, Sand 1, 72076 T\"ubingen, Germany}

\def\LaTeX{L\kern-.36em\raise.3ex\hbox{a}\kern-.15em
    T\kern-.1667em\lower.7ex\hbox{E}\kern-.125emX}

\begin{document}

\label{firstpage}

\maketitle

\begin{abstract}

We present a 3D photoionization model of the planetary nebula NGC~6302, one of the most complex and enigmatic objects of its kind. It's highly bipolar geometry and dense massive disk, coupled with the very wide range of ions present, from neutral species up to Si$^{8+}$, makes it one of the ultimate challenges to nebular photoionization modelling.

Our {\sc mocassin} model is composed of an extremely dense geometrically thin circumstellar disk and a large pair of diffuse bipolar lobes, a combination which was necessary to reproduce the observed emission-line spectrum. The masses of these components, 2.2~M$_\odot$ and 2.5~M$_\odot$ respectively, gives a total nebular mass of 4.7~M$_\odot$, of which 1.8~M$_\odot$ (39\%) is ionized. Discrepancies between our model fit and the observations are attributed to complex density inhomogeneities in the nebula. The potential to resolve such discrepancies with more complex models is confirmed by exploring a range of models introducing small-scale structures. Compared to solar abundances helium is enhanced by 50\%, carbon is slightly subsolar, oxygen is solar, and nitrogen is enhanced by a factor of 6. These all imply a significant 3rd dredge-up coupled with hot-bottom burning CN-cycle conversion of dredged-up carbon to nitrogen. Aluminium is also depleted by a factor of 100, consistent with depletion by dust grains.

The central star of NGC~6302 is partly obscured by the opaque circumstellar disk, which is seen almost edge-on, and as such its properties are not well constrained. However, emission from a number of high-ionization `coronal' lines provides a strong constraint on the form of the high-energy ionizing flux. We model emission from the central star using a series of stellar model atmospheres, the properties of which are constrained from fits to the high-ionization nebular emission lines. Using a solar abundance stellar atmosphere we are unable to fit all of the observed line fluxes, but a substantially better fit was obtained using a 220,000~K hydrogen-deficient stellar atmosphere with $\log g = 7.0$ and $L_\star = 14,300$~L$_\odot$. The H-deficient nature of the central star atmosphere suggests that it has undergone some sort of late thermal pulse, and fits to evolutionary tracks imply a central star mass of 0.73--0.82~M$_\odot$. Timescales for these evolutionary tracks suggest the object left the top of the asymptotic giant branch $\sim$2100 years ago, in good agreement with studies of the recent mass-loss event that formed one pair of the bipolar lobes. Based on the modelled nebular mass and central star mass we estimate the initial mass of the central star to be 5.5~M$_\odot$, in approximate agreement with that derived from evolutionary tracks.

\end{abstract}

\begin{keywords}
planetary nebulae: individual: NGC 6302, ISM: abundances
\end{keywords}

\section{Introduction}

Planetary nebulae (PNe) are one of the last evolutionary stages of the majority of low and intermediate mass stars (1-8~M$_{\odot}$). During the previous asymptotic giant branch (AGB) evolutionary phase the star ejects a large fraction of its mass during brief phases ($10^3 - 10^4$~yrs) of high mass-loss (up to $10^{-4}$~M$_{\odot}$~yr$^{-1}$). This mass-loss eventually exposes the central core of the star that radiates strongly in the UV and ionizes the surrounding nebula making it visible for study. Despite supposed spherically symmetric mass-loss while on the AGB, many PNe show point- or axi-symmetric shapes and complex morphologies that have been interpreted as evidence for the influence of a binary companion \citep[e.g.][]{dema09}. This and other outstanding issues in PNe research such as their small-scale structures \citep[e.g.][]{gonc01,mats09} and discrepancies between forbidden line and recombination line abundances for heavy elements \citep{liu00,tsam04,wess05} suggest that out understanding of the late stages of stellar evolution is far from complete. Despite this, PNe are still regularly used as standard candles, metallicity indicators and tracers of stellar populations in distant galaxies \cite[e.g.][]{buzz06}. If PNe are to be used in this way our understanding of their structure and evolution must be greatly improved, both through studies of large samples of such objects \citep[e.g.][]{viir09,sabi10}, their precursors the AGB stars \citep[e.g.][]{vand88,zijl01,wrig08}, and through individual case studies to understand the most complex examples \citep[e.g.][]{gonc06,ware06}.

\begin{figure}
\begin{center}
\includegraphics[width=180pt]{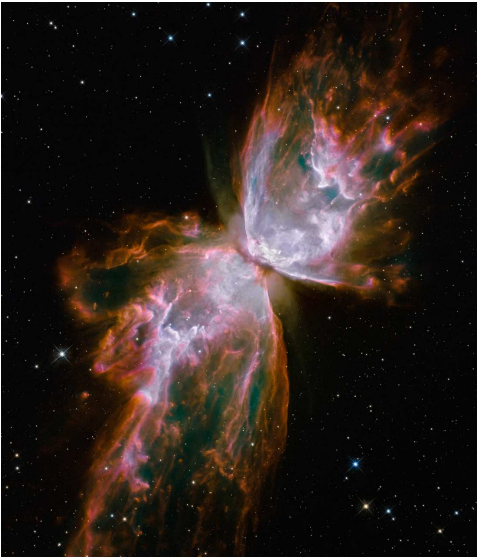}
\caption{The planetary nebula NGC~6302 as imaged with the Wide Field Camera on the Hubble Space Telescope (HST). Blue is He~{\sc ii}, brown is H$\alpha$, and white is [S~{\sc ii}]. The image is approximately $2 \times 2.5$~arcmin or $0.6 \times 0.75$~pc at the distance of NGC~6302. Image copyright: NASA and ESA.}
\label{ngc6302}
\end{center}
\end{figure}

NGC~6302 is one of the most complex and extreme examples of a planetary nebula. The nebula is broadly bipolar, with a second narrower pair of lobes visible further out (see Figure~\ref{ngc6302}). It has a highly pinched waist that is typical of `butterfly' bipolar PNe \citep{bali02}, as is a highly complex structure with many clumps and knots. The bipolar structure has been attributed to the presence of a very dense circumstellar disk that is seen almost edge-on and almost completely obscures the central star \citep[$A_V \sim 6.6$,][]{mats05,szys09}, which has only recently been detected. The disk radiates strongly in the infrared \citep[e.g.][]{lest84} with evidence for significant fractions of very large grains \citep{hoar92} and crystalline silicates \citep{mols01}. NGC~6302 has been classified as a Type~I PN, both according to the criteria or \citet[][N/O $\geq$ 0.5 {\em and} He/H $\geq$ 0.125]{peim83} and the criteria of \citet{king94} for a Milky-Way Type~I PN, that N/O $\geq$ 0.8. The high helium and nitrogen abundances observed in the nebula imply a potentially massive progenitor, while observations of emission lines from very high ionization stages \citep[e.g.][]{ashl88,rowl94,casa00} suggest an extremely hot central star.

Attempts to understand NGC~6302 have been hindered by its complexity and the inability to study its central star and therefore determine its evolutionary history. \citet{alle81} could find ``no uniform density theoretical model that could reproduce the line intensities in NGC~6302 in any satisfactory fashion", while \citet{lame91} could not produce a physically plausible model for the nebula without requiring shock-excitation, which \citet{orig93} and \citet{casa00} found strong evidence against. However, many of these models were forced to assume spherical symmetry due to the computational constraints of full 3-dimensional modeling, yet the inhomogeneous morphology of NGC~6302 clearly requires a 3-dimensional model if reliable conclusions are to be drawn.

In this paper we present a 3D photoionization model of NGC~6302 using the 3D Monte Carlo photoionization and radiative transfer code {\sc mocassin} \citep{erco03,erco05,erco08}. The model presented here is a fully 3-dimensional model of NGC~6302 that includes both gas and dust, though we will only discuss the gas photoionization model in this paper. We will present the dust radiative transfer model in a future paper. In Section~2 we discuss the changes made to the {\sc mocassin} code to enable the modeling of this object and outline the observations that our models are to reproduce. In Section~3 we describe the model used and the parameters used to build the model. In Section~4 we describe the modeling process and how different quantities varied according to changes in different parameters. In Section~5 we present the model results, including fits to the observed emission-line spectrum and the ionization and temperature structure of the nebula. Finally, in Section~6 we describe the resulting nebula properties and discuss the evolutionary history of the object.

\section{The {\sc mocassin} code}

{\sc mocassin} is a 3D photoionization and radiative transfer code that uses Monte Carlo methods to solve the radiative transfer in a photoionized nebula \citep{erco03}. The Monte Carlo approach allows it to fully treat all physical processes in 3D, self-consistently treating the diffuse component of the radiation field that 1D photoionization codes are unable to deal with. The code is highly versatile and can model any structure or morphology, introducing unlimited subgrids to resolve detail in areas of the model. It also allows multiple non-central ionizing sources and multiple gas chemistries. {\sc mocassin} also has a complete radiative transfer dust model including the full treatment of all dust processes such as photoelectric heating and gas-grain collisions \citep{erco05}. The atomic database included opacity data from \citet{vern93} and \citet{vern95}, energy levels, collisions strengths and transition probabilities from Version~5.2 of the CHIANTI\footnote{http://www.ukssdc.ac.uk/solar/chianti/} \citep[][and references therein]{dere97,land06} and the improved hydrogen and helium free-bound continuous emission data of \citet{erco06}. In this section we describe a number of changes made to the code as part of this work, as well as details of the computational facilities that the models were run on, and the observations that were to be reproduced by the models.

\subsection{2-dimensional simulations in {\sc mocassin} }

{\sc mocassin} is a 3-dimensional code, but many objects exhibit cylindrical symmetry along a rotation axis. Such objects may effectively be modeled in two dimensions, $r$ and $z$, greatly reducing the computational and memory requirements of large and complicated models. We have developed and tested a 2-dimensional version of the {\sc mocassin} code, which integrates the observations over $\theta$ from 0 to 360 (where $\theta$ is the angle between two directions in the $z=0$ plane). Since the nebula is symmetrical upon reflection in the $z = 0$ plane, only positive $z$ values need be modeled, reducing the processing time further. This approximation still properly treats the diffuse component of the radiation field and thorough benchmarking has been performed that fully reproduces all of the {\sc mocassin} benchmarks described in \citet{erco03}. Although the models described here were run in 2D, we will often discuss the resulting 3D structure of the nebula, since the code reproduces the full 3D structure in the outputs. The final model was run in 3D to confirm the validity of the 2D modeling approach and no differences were found. The new routines are included in the latest public version of the code \citep{erco08}.

\subsection{Computational details}

The models presented in this work were run on the {\sc hiperspace} facilities at University College London, including the {\it Enigma} SUN Microsystems V880 multiprocessor computer, the {\it Keter} SUN Microsystems V880 and V890 microprocessor computers and the Dell {\it Legion} Cluster, which uses Intel's dual-core technology. Simulations were typically run with 8-16 processors and approximately 32~GB of memory. The necessary computational processing time varied depending on the complexity of the model and the required level of convergence of the Monte Carlo simulation, typically an iteration-to-iteration variation of the fraction of neutral hydrogen of $<$1\% in $>$60\% of the cells in the model.

\subsection{Observations to reproduce}
\label{s-observations}

{\sc mocassin} simulations produce several outputs that are used to compare the model results with observations. The outputs can also be adapted to the exact setup of the observations by simulating the slit or aperture through which the observations were made and the inclination angle of the nebula to the line of sight. As shown by \citet{gonc06} this is particularly important for long-slit observations of highly inhomogeneous nebulae such as NGC~7009 and the broad range of different measurements for NGC~6302 of the same emission line is testament to this. We have searched the literature for observations of NGC~6302 that span all wavelength regimes and that include lines from all the important elements and their ionization stages. The list in Table~\ref{ngc6302_specobs} was chosen based on spectral coverage and resolution. Where measurements overlap the most precise measurement was taken, or that which was accompanied by complementary information such as electron temperatures or densities derived from diagnostic line ratios.

\begin{table*}
\footnotesize
\begin{center}
\caption{Spectroscopic observations to be matched by the {\sc mocassin} photoionisation model of NGC~6302.}
\begin{tabular}{@{}lllccl}
\hline
Region & Wavelength range & Telescope & Lines & Aperture / slit & Reference \\
\hline
Ultraviolet & 1150-1975 \AA\ & IUE SWP & 7 & 10.3\arcs $\times$ 23\arcs (oval) & \citet{tsam03} \\
Ultraviolet & 3426 \AA\ & Mt. Lemmon 1.5-m & 1 & 48\arcs (circular) & \citet{rowl94} \\
Optical & 3040-7400 \AA\ & ESO 1.52-m & 30 & 210\arcs $\times$ 2\arcs & \citet{tsam03} \\
Optical-IR & 3300-8600 \AA\ & Siding Springs 2.3-m & 1 & 2\arcs (long-slit) & \citet{grov02} \\
Optical-IR & 3500-20,000 \AA\ & Cerro Tololo 92-cm & 1 & 17\arcs $\times$ 34\arcs & \citet{danz73} \\
Near-infrared & 2.7-4.8 $\mu$m & UKIRT (echelle) & 12 & 3\arcs $\times$ 3\arcs & \citet{casa00} \\
Mid-infrared & 2.4-45 $\mu$m & ISO SWS & 14 & 14\arcs $\times$ 20\arcs & \citet{bein99} \\
Mid-infrared & 8-13 $\mu$m & UKIRT N-band & 5 & 4\arcs (circular) & \citet{casa00} \\
Far-infrared & 43-198 $\mu$m & ISO LWS & 7 & 40\arcs (circular) &\citet{liu01} \\
\hline
\end{tabular}
\label{ngc6302_specobs}
\end{center}
\end{table*}

The UV spectrum is mainly taken from the International Ultraviolet Explorer (IUE) spectrum presented by \citet{tsam03} and complemented by observations of the [Ne~{\sc v}]~3426~\AA\ line by \citet{rowl94} for comparison with the [Ne~{\sc v}] lines in the mid-IR. The main optical emission-line spectrum was that of \citet{tsam03}, a deep optical spectrum made with a fixed slit aligned along the polar axis of the nebula that covers all aspects of the nebula. These observations include many of the diagnostic line ratios that will be important to diagnose the density and ionization structure within the nebula. This main line list is supplemented by lines from the far-red region using the published spectra of \citet{danz73} and \citet{grov02}. The infrared line list is split between the {\it Infrared Space Observatory} (ISO) observations using the Short Wavelength Spectrometer (SWS) presented by \citet{bein99} and the Long Wavelength Spectrometer (LWS) from \citet{liu01}, and complemented by deep echelle spectroscopy from \citet{casa00}. These include many fine-structure lines in the mid- and far-infrared which have the advantage of only having a weak temperature dependence due to their low excitation energies. 

In cases where the observed lines are known to be blends that are inseparable in the observations (e.g. the blend of [Ar~{\sc iv}]~$\lambda$4711 and [Ne~{\sc iv}]~$\lambda$4712), we accounted for the presence of the undesired component by calculating its intensity relative to other lines of the same species using the {\sc equib} code.

Where possible all observations have been converted onto a scale relative to H$\beta$ for comparison with modeled line fluxes. For observations where this is not possible we compare observed fluxes, calculating model line fluxes using the distance to NGC~6302 of 1.17~kpc measured by \citet{meab08}. We also compare the H$\beta$ flux for the entire nebula from our models with that measured by \citet{tsam03} as F(H$\beta) = 2.82 \times 10^{-11}$~erg~cm$^{-2}$~s$^{-1}$. Using their derived reddening for NGC~6302 (c(H$\beta) = 1.44$) we obtain a dereddened H$\beta$ flux of $7.76 \times 10^{-10}$~ergs~cm$^{-2}$~s$^{-1}$ or L($\beta) = 1.27 \times 10^{35}$~ergs~s$^{-1}$, which can be compared with model outputs for the nebula.

\subsection{Observational constraints on the central star luminosity}
\label{s-starlum}

The central star of NGC~6302 was identified for the first time in deep {\it Hubble Space Telescope} images reported by \citet{szys09}, who were able to perform photometric measurements of the central star in two filters and therefore make estimates of the surface temperature, extinction and stellar luminosity. However, at such high temperatures and extinctions, the exact solution is not well constrained and the authors were forced to use constraints on the age of the nebula, combined with evolutionary models, to fit the blackbody properties of the central star as (T,L)~=~(200,000~K,~2000~L$_\odot$). However, uncertainties on the photometry, evolutionary models, and extinction law led the authors to derive luminosities ranging from 860 -- 24,000~L$_{\odot}$ depending on these assumptions.

Another method for estimating the luminosity of the central ionizing source is based the sum of all emitted and observed radiation from the nebula. To perform this sum we have combined observations from a number of different observatories and wavelengths, the results of which are listed in Table~\ref{cs_luminosity}. The IUE observations, which are for a $23 \times 10$ arcsec aperture, were scaled up to the whole nebula using the intrinsic He~{\sc ii} $\lambda$1640/$\lambda$4686 ratio \citep[$\sim$7 for $T \sim 15000$~K, $N_e \sim 10^5$~cm$^{-3}$, Case B,][]{oste06} and the dereddened $\lambda$4686 line flux from \citet{tsam03}. The UV -- radio continuum emission is primarily free-free and bound-free emission, the former of which was estimated using the equations of \citet{alle77}, while the latter was calculated using the 2-photon continuum tool {\tt nebcont} within {\sc dipso}\footnote{http://star-www.rl.ac.uk/docs/sun50.htx/sun50.html}. The infrared SED was taken from the ISO SWS and LWS spectra, with the ISO SWS spectrum was scaled up by a factor of 1.45 to match the LWS spectrum.

\begin{table*}
\footnotesize
\begin{center}
\caption{Total luminosity of NGC~6302 derived from all (dereddened) observed and theoretical components, assuming a distance of 1.17~kpc \citep{meab08}. See text for a discussion of the individual methods used.}
\begin{tabular}{@{}lllr}
\hline
Component & Source & Reference & Luminosity / $L_{\odot}$ \\
\hline
Ultraviolet line fluxes & IUE & \citet{feib01} & 104 \\
Optical line fluxes & ESO 1.52-m & \citet{tsam03} & 1324 \\
Far-red line fluxes & Siding Springs 2.3-m & \citet{grov02} & 73 \\
Near-IR line fluxes & UKIRT & \citet{casa00} & 140 \\
Free-free continuum & Theoretical & \citet{alle77} & 122 \\
Bound-free continuum & Theoretical & {\sc dipso} & 837 \\
Infrared continuum & ISO & \citet{mols01} & 3090 \\
\hline
Total & & & 5690 \\
\hline
\end{tabular}
\label{cs_luminosity}
\end{center}
\end{table*}

The total luminosity of all components is $\sim$5700~L$_{\odot}$. This is likely to be a lower limit for the luminosity of the central star because the nebula and circumstellar disk are seen edge on and therefore a fraction of radiation from the central star may be escaping along the polar axis. Based on the geometry of the model nebula we have used, the opening angle of the circumstellar disk, and the reprocessing of radiation in the bipolar lobes, we estimate that the fraction of radiation escaping along the polar axis may be as much as a factor of two. We therefore estimate the luminosity of the central star to be $\sim 5700 - 11,400$~L$_{\odot}$. This value is higher than previous estimates, which have either sampled only one component of the nebula \citep[e.g.][estimated a nebular luminosity of 3300~L$_{\odot}$ based on infrared dust modeling]{mats05} or have been estimated from evolutionary models \citep[][estimated 2000~L$_{\odot}$]{szys09}.

\section{The model nebula}

The final model described in this paper was built up over the course of this work starting with a simple pair of bipolar lobes and growing in complexity as both the models demanded and more observations were added to the list of observations to match. Every part of this model was necessary in some way to reproduce the observations described above. The principle components of the model are a large pair of bipolar lobes and a very dense disk-like distribution of circumstellar material. In this section we describe the different components of the NGC~6302 model and how this was constructed using the {\sc mocassin} photoionization code.

Though observations show the presence of two pairs of bipolar lobes around NGC~6302 \citep[e.g.][]{meab05}, we modeled only one pair for simplicity. Their similar orientations and the fact that most slits only cover the inner regions of the nebula suggest this will not significantly affect our results. The presence of large amounts of circumstellar material has been inferred from observations of a dark lane obscuring the central region of the nebula \citep{mats05} and observations of molecular material in the core of NGC~6302 \citep{pere07}. We attempted to model this material as both a circumstellar disk or a torus,  the difference being that a torus has a scale height similar to its radial scale length, while a disk is intrinsically flat with a scale length significantly larger than its scale height. We found that the observations were better fit by a circumstellar disk (this will be discussed further in Section~\ref{discuss-structure}).

These structures were all modelled as cylindrically symmetric with the same axes as that of the bipolar lobes. This was a necessary simplification for this complicated nebula, but it should be noted that radio observations suggest that the disk is tilted with respect to the lobes, and most-likely warped \citep{gome89,mats05}. All these components have a number of parameters that were initially estimated from previous observations of NGC~6302 but were allowed to vary to obtain the best fit during the modeling process. The final set of parameters used, and their relative uncertainties, are provided in Section~\ref{discuss-model}.

\subsection{The bipolar lobes}
\label{s-model}

\begin{figure}
\begin{center}
\includegraphics[width=235pt]{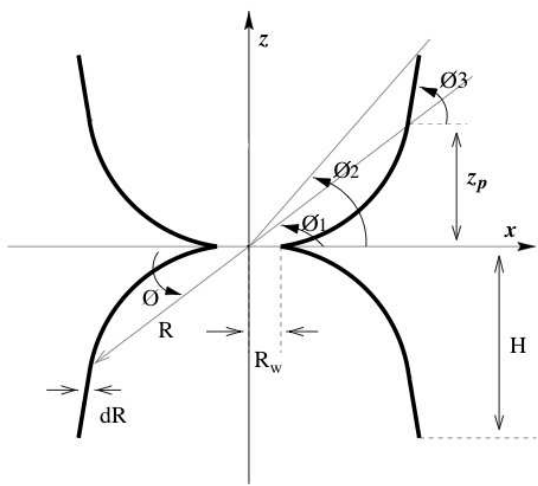}
\caption{Schematic view of the geometrical structure of an unfilled hourglass-shaped bipolar nebula with rotational symmetry about the $z$-axis. See text for an outline of the parameters governing the structure and Table~\ref{parameters} for a list of values used in the model of NGC~6302. Figure adapted from one by \citet{daya00} for their model of MyCn18.}
\label{structure6302}
\end{center}
\end{figure}

The bipolar lobes used in our model were developed from the model described by \citet{daya00} for their model of MyCn~18. This geometrical model, shown in Figure~\ref{structure6302}, is based around a truncated paraboloid near the central star that develops into a cone-like structure further away from the star. The parameters indicated in Figure~\ref{structure6302} are the latitude, $\phi_1$, of transition from the paraboloid to the cone, the latitude, $\phi_2$, of the end of the cone, the slope, $\phi_3$, of the cone, the semi-height, $H$, of the hourglass, and the waist-radius, $R_w$, of the paraboloid. The shape of the walls of the hourglass can then be described using the equations

{\setlength\arraycolsep{5pt}
\begin{eqnarray}
 & x^2 + y^2 = \left( \frac{\displaystyle z_p}{\displaystyle \mathrm{tan}^2 \phi_1} - \frac{\displaystyle R_w^2}{\displaystyle z_p} \right) z + R_w^2 & | z | \le z_p \\
 & \sqrt{x^2 + y^2} = \frac{\displaystyle z}{\displaystyle \mathrm{tan} \phi_3} - z_p \left( \frac{\displaystyle 1}{\displaystyle \mathrm{tan} \phi_3} - \frac{\displaystyle 1}{\displaystyle \mathrm{tan} \phi_1} \right)	& | z | > z_p
\end{eqnarray}
}

\noindent where the transition height, $z_p$, is given by

\begin{equation}
z_p = H \frac{ ( \mathrm{tan} \phi_3 / \mathrm{tan} \phi_2 ) - 1 }{ (\mathrm{tan} \phi_3 / \mathrm{tan} \phi_1 ) - 1}
\end{equation}

\noindent and the radial distance from the hourglass center, $R$, and the latitude, $\phi$, of a point in the nebula are given by

{\setlength\arraycolsep{5pt}
\begin{eqnarray}
R = (x^2 + y^2 + z^2)^{1/2} & &  \phi = \mathrm{tan}^{-1} \left( \frac{z}{(x^2 + y^2)^{1/2}} \right) .
\end{eqnarray}
}

\begin{figure}
\begin{center}
\includegraphics[width=235pt]{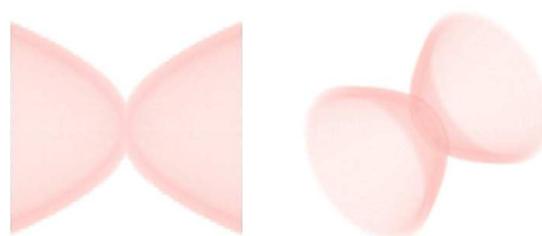}
\caption{Visualisation of the three-dimensional density structure for the bipolar lobes of NGC~6302 using the model illustrated in Figure~\ref{structure6302} and using the parameters listed in Table~\ref{parameters}. {\it Left:} Side-on image, showing the hourglass structure. {\it Right:} Image of the bipolar lobes inclined at an angle of 45$^{\circ}$.}
\label{lobes_visual}
\end{center}
\end{figure}

The dimensions of the model nebula were estimated from images of NGC~6302 such as that in Figure~\ref{ngc6302} combined with an adopted inclination angle of $12.8\,^{\circ}$ \citep{meab05} and distance of $1.17 \pm 0.14$~kpc \citep{meab08}. We estimated the height of the lobes to be $H = 7.0 \times 10^{17}$~cm, the latitudes of the parabola and cone as $\phi_1 = 41.5\,^{\circ}$, $\phi_2 = 49.6\,^{\circ}$ and the slope of the cone as $\phi_3 = 66.0\,^{\circ}$. These quantities were all constrained by images of the nebula and were not varied during the modeling process. Estimated measurement uncertainties are $\pm 10$\% for $H$, $\pm 25$\% $\phi_1$, and $\pm 10$\% for $\phi_2$ and $\phi_3$. The waist radius, $R_w$, was set to be the same value as the inner radius of the circumstellar disk and could be varied during the modeling process. Using these parameters the resulting nebula is quite wide (see Figure~\ref{lobes_visual}) as is seen in observations of NGC~6302. Finally, the density and density gradient in the lobes, as well as whether the lobes should be hollow (i.e. only edge-brightened) or filled was allowed to vary during the modeling process.

\subsection{The circumstellar disk}
\label{s-disk}

\begin{figure}
\begin{center}
\includegraphics[width=235pt]{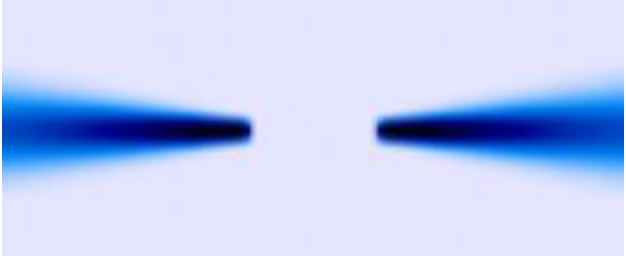}
\caption{Visualisation of the density distribution in the circumstellar disk seen as a slice through the $x$-$z$ plane. the image covers a region of $1 \times 10^{17}$ by $4 \times 10^{16}$~cm. The density is represented logarithmically in blue and covers the range $10^4$~-~$10^6$~cm$^{-3}$. Note the flaring in the outer disk.}
\label{disk_visual}
\end{center}
\end{figure}

To model the circumstellar disk we used the flared Keplerian disk model of \citet{pasc04}. In their model, the density distribution of the disk has the form

\begin{equation}
N_H (r, z) = N_0 \left( \frac{r}{r_d} \right)^{-\alpha} \mathrm{exp} \left( \frac{-0.25 \, \pi \, z^2}{(z_d (r / r_d)^{f_d} )^2 } \right)
\label{disk_equation}
\end{equation}

\noindent where $r$ is the radial distance and $z$ is the distance from the midplane. $r_d$ and $z_d$ are disk length and height parameters, which determine the scale length of the density distribution, and $N_0$ is a characteristic density for the disk. The radial density dependence of $\alpha = 1$ used by \citet{pasc04} is also that found by \citet{mats05} from a best-fit to the spectral energy distribution (SED) in their radiative transfer models, though we allowed this to vary in our models. The flaring of the disk with distance from the central star is given by $z_d (r / r_d)^{f_d}$, utilising the flaring parameter $f_d$, which controls the opening angle as a function of distance from the central star (see Figure~\ref{disk_visual}). The disk extends from an inner radius, $r_{in}$, to an outer radius, $r_{out}$. All these parameters were allowed to vary in an attempt to find the best fitting model.

Initial parameters for the circumstellar disk were taken from the SED fitting of \citet{mats05} such as the height parameter, $z_d = 5.0 \times 10^{15}$~cm, and the inner radius, $r_{in} = 1.0 \times 10^{16}$~cm. For other parameters we used the initial estimates given by \citet{pasc04}. The outer radius was initially set as $r_{out} = 1.0 \times 10^{17}$~cm. The characteristic density, $N_0$, was also allowed to vary during the modeling process.

\subsection{The model grid}

The bipolar lobes and circumstellar disk were mapped onto a 2D grid in the $x$-$z$ plane, utilising the fact that the model nebula exhibits both cylindrical symmetry and is symmetric upon reflection in the $z=0$ plane (though, as noted above, the observed nebula is not cylindrically symmetric and is better described as point-symmetric). The number of cells in the grid and the size of the cells was varied during the modeling process to ensure that the ionization fronts and temperature and density gradients in the model were appropriately resolved such that the resulting emission line spectrum was a true representation of the model nebula. The final model consisted of a 2D grid of $138 \times 86$ cells with the cell width varying from $2 \times 10^{14}$ in the circumstellar disk, to $5 \times 10^{16}$ in the outer regions of the bipolar lobes. This was then mapped onto a 3D grid of $138 \times 138 \times 86$ cells to confirm the validity of the 2D models and test the influence of density inhomogeneities in 3D (discussed below).

\subsection{The central star}
\label{s-centralstar}

The central ionizing source of NGC~6302 was modeled using stellar model atmospheres produced using the {\it T\"ubingen NLTE Model Atmosphere Package}\footnote{Website: http://astro.uni-tuebingen.de/\~rauch/TMAP/TMAP.html. In the framework of the Virtual Observatory (\emph{VO}, http://ww.ivoa.net), all spectral energy distributions from the \emph{TMAP} model grids described here are available in \emph{VO} compliant form from the \emph{VO} service \emph{TheoSSA} (http://vo.ari.uni-heidelberg.de/ssatr-0.01/TrSpectra.jsp?) provided by the \emph{German Astrophysical Virtual Observatory} (\emph{GAVO}, http://www.g-vo.org).} \citep[TMAP,][]{rauc03b,wern03}. While it is sometimes true that the assumption of a blackbody is not worse than any other assumption in producing a photoionization model fit, at energies higher than 13.6~eV and 54.4~eV the differences between a blackbody flux and a stellar atmosphere can become so great that very large differences will result \citep[e.g.][]{rauc97,rauc03,arms02}. A grid of stellar atmosphere models were tested in our simulations, with effective temperatures ranging from 50,000 to 300,000~K and $\log g$ from 5-9. These were tested for both solar ($[X] = [Y] = [Z] = 0$) and metal-poor ($[X] = [Y] = 0$, $[Z] = -1$) elemental abundances \citep[where the solar abundances used by TMAP are those from][and include all the elements up to nickel]{aspl05}. Model atmosphere fluxes were also tried for `typical' PG~1159 abundance ratios (He:C:N:O~=~33:50:2:15 by mass), covering the same effective temperature and gravity ranges. Given the lack of knowledge about the central star of NGC~6302 we allowed all these properties to vary in our models.

\subsection{Nebular abundances}
\label{s-origabundances}

No spatially resolved spectroscopic observations of NGC~6302 have been published, so there is no direct observational evidence for structures within the nebula with different elemental abundances. For this reason we used a homogeneous elemental abundance distribution in our models, despite {\sc mocassin}'s ability to model regions with different chemistries. 
Our model used 14 elements, including all the principle elements that contribute to the heating and cooling of the nebula, as well as those responsible for the important density and temperature sensitive line ratios, and those required to reproduce the high-ionization coronal lines observed by \citet{casa00}. The initial abundances of He, C, N, O, Ne, S, Cl and Ar were taken from \citet{tsam03}, the abundances of Mg and Al were taken from the analysis of \citet{casa00}, while abundances for Na, Si and K were initially set to solar values \citep{lodd03}. All abundances were initially kept constant while structural parameters were being allowed to vary, but were then varied to obtain the best fit to the lines from each element (see final values in Table~\ref{parameters}).

\subsection{Dust modeling}

NGC~6302 is known to have a very large dust component \citep[e.g.][]{mols01,kemp02b} that will affect the radiative transport in the nebula. Any photoionization model of the nebula must therefore fully consider the influence of dust on the ionization structure of the gas. The models presented here therefore include a dust component that utilises {\sc mocassin}'s full treatment of dust radiative transfer \citep[e.g.][]{erco05}.The dust model will be fully discussed in a future paper, but we outline here the essential properties. Dust was included in the form of a number of dust species whose size distribution was adapted from the standard MRN model \citep{math77}. Optical constants were obtained from the literature and converted to absorption efficiencies using {\sc mocassin}'s light scattering codes (that treat both spherical and elliptical dust grains). There is limited availability of UV and far-UV optical constants for common dust species in the literature, the most widely used being the amorphous silicates of \citet{drai03b}, which were adopted in our model. Dust absorption within the nebula is significant in reducing the ionizing flux, with approximately half of all Lyman photons absorbed by the dust (in agreement with the observed fraction of luminosity coming from the IR SED, see Section~\ref{s-starlum}).

\section{The modeling process}

In this section we describe the modeling process that was used to constrain the model properties by fitting the different observations. In particular we describe how different model parameters caused variations in the observable lines intensities or line ratios and how this led to the final model. These dependencies of lines or ratios on certain parameters were not absolute and changing other parameters could sometimes nullify the influence of certain parameters. This made the modeling process highly complicated and often required the parameter space around a model to be independently investigated at regular intervals in case new parameter dependencies had emerged due to changes in the model. Furthermore, in a model as complex as this there will inevitably be degeneracies in the models that can fit the observations, and some lines or ratios may remain not fitted. The objective here is to highlight which parts of the model are the most constrained by the observations and therefore represent the most important features of the model. The final model, while not a perfect fit, avoids many of the poor fits that certain parameters resulted in. We have divided this discussion, for the purposes of clarity, into the fundamental areas of the model (nebular structure, central star properties and nebula chemistry), but it should be noted that all the parameters were fitted simultaneously.

\subsection{The nebular geometry and density}
\label{s-geometry}

The geometry and density of the nebula, specifically its two main components, the circumstellar disk and the bipolar lobes, were refined by fitting the observed density- and temperature-sensitive line ratios. These line ratios respond to changes in the density and ionization structure of the nebula, but also highlight specific regions of the nebula according to their ionization potentials. The diagnostic line ratios can therefore be divided into those primarily responsive to changes in the low-density bipolar lobes, or the high-density circumstellar disk. These lists were not independent and many line ratios, particularly those sensitive to changes in the ionization structure of the nebula, were influenced by a large range of parameters. This was further complicated by the apertures and slits of the different observations of NGC~6302, each of which sampled a mixture of the two structural components. Despite this, many line ratios very highly responsive to specific parameters and by fitting these line ratios many structural parameters could be constrained. With these restrictions incorporated the variation of the remaining parameters was explored and the best-fitting model identified.

The circumstellar disk is modeled with seven parameters ($N_0$, $r_{in}$, $r_{out}$, $r_d$, $\alpha$, $z_d$, $f_d$), though the characteristic density, $N_0$, and the inner radius, $r_{in}$, were the most influential in the resulting emission line spectrum and are therefore the most well constrained. This is because these parameters influence the region of the disk exposed to the full ionizing flux, while many of the other parameters affect regions of the disk beyond the ionization front. The line ratios that were responsive to changes in the circumstellar disk are a mixture of those that probe either the ionization structure, or regions of high temperature or density.

The most influential parameter was the characteristic density of the circumstellar disk, $N_0$, which we fit to a value of $N_0 = 80,000$~cm$^{-3}$, with an accuracy of approximately $\pm$20\%. This parameter was influenced by nearly all the observed line ratios, though the strongest constraint came from the He~{\sc ii} / He~{\sc i} line ratio which was over-estimated by a factor 10 if $N_0 < 60,000$~cm$^{-3}$, a disparity that could not be resolved by adjusting any other parameters. Many of the density-sensitive line ratios with high critical densities (i.e. those of [Cl~{\sc iii}], [Ar~{\sc iv}], and [K~{\sc v}] ) could also only be fit with a very dense circumstellar disk. Very little variation was observed in any of the line ratios when different density profiles were adopted, primarily because the circumstellar disk is highly optically thick. The initial choice of $N_H \propto r^{-1}$ for the circumstellar disk radial density profile was made on the basis of the SED fitting of \citet{mats05}, but $r^{-1/2}$ or $r^{-2}$ density profiles led to little difference.

The inner radius of the circumstellar disk, $r_{in}$, was also very influential as it dictates the strength of the ionizing flux at the circumstellar disk. It was best constrained by a mixture of the [O~{\sc ii}] (3726 + 3729) / (7320 + 7330) and [Ar~{\sc iv}] density-sensitive line ratios. The former was over-estimated by a factor 5 if $r_{in} > 1.4 \times 10^{16}$~cm, and the latter increased significantly as $r_{in}$ was decreased. The final value of $r_{in} = 1.2 \times 10^{16}$~cm was chosen to balance these two ratios though it could be decreased if the [Ar~{\sc iv}] line ratio could be reduced by other methods. At a distance of 1.17~kpc this is equivalent to an angular radius of 0.68$^{\prime\prime}$.

The thickness of the circumstellar disk (determined by $z_d$) influenced a large number of line ratios, most critically the [O~{\sc ii}] 3729 / 3726 ratio which became irretrievably over-estimated if the disk thickness was increased, and the [N~{\sc ii}] 5755 / (6584 + 6548) ratio that greatly increased if the thickness was significantly decreased. The best fit value of $z_d = 6.0 \times 10^{15}$~cm was constrained to an uncertainty of $\pm$50\%.

The other disk parameters only influence material beyond the ionization front and therefore had a lot less influence on the modeled spectrum. The flaring parameter, $f_d$, was only constrained to be $\geq 1.1$ by the [O~{\sc iii}] / [O~{\sc ii}] and [S~{\sc ii}] 4068 / (6731 + 6717) line ratios (which both increased significantly if the parameter was reduced), but no upper constraints were identified. It was therefore kept at its initial value, $f_d = 1.125$. The length parameter, $r_d$, and the outer radius, $r_{out}$, influenced a small number of line ratios, but none sufficiently enough to constrain them. The former parameter was therefore kept at its initial value of $r_d = 5.0 \times 10^{16}$~cm, while the latter parameter was adjusted based only on the results of radiative transfer dust-fitting (not discussed in this paper) and therefore we list its value as unconstrained here.

The circumstellar material was modeled here as a flat disk, but attempts were made to model it as a torus (i.e. a circumstellar distribution with a greater height and smaller length), as suggested by \citet{mats05} and \citet{pere07}. However, increasing the disk thickness to that suggested by \citet{pere07} resulted in an unfittable [O~{\sc ii}] 3729 / 3726 ratio (as noted above). We therefore suggest that certainly in the inner regions where the ionized gas traces the structure of the circumstellar material, it is in the form of a thin disk. In the outer, neutral regions the circumstellar material may better represent a torus if traced by neutral material or dust.

The shape and dimensions of the bipolar lobes were adopted from images of the nebula such as that in Figure~\ref{ngc6302} with the distance to the nebula from \citet{meab08}. These parameters were fixed throughout the modeling process and only the density of the lobes and the density profile were varied. The lobes were initially modeled with a low density of 2000~cm$^{-3}$ constrained by a number of line ratios that are particularly sensitive to density such as the [O~{\sc ii}] 3729~/~3726, [O~{\sc iii}] 4363 / 5007, [O~{\sc iii}] 52~/~88~$\mu$m, and [S~{\sc ii}] ($\lambda$4068 + $\lambda$4076) / ($\lambda$6731 + $\lambda$6717) ratios (all sensitive to densities $< 5000$~cm$^{-3}$). However, none of these line ratios offered a particularly strong constraint on the final density, with the [O~{\sc ii}] 3729~/~3726 line ratio requiring densities in excess of 1000~cm$^{-3}$, while the [O~{\sc iii}] 52~/~88~$\mu$m line ratio increased if the density was significantly increased above 4000~cm$^{-3}$. A final uncertainty of $\pm$50\% is estimated for the final value of 2000~cm$^{-3}$ based on these constraints. No constraint could be found on the density profile of the bipolar lobes so they were kept at a constant density.

In response to an initially poor fit to a number of density-sensitive line ratios such as those of [Ar~{\sc iv}] and [K~{\sc v}], as well as the temperature-sensitive [N~{\sc ii}] line ratio, a third component was added to our model. This region, dubbed `the outflow', is an inner region of the bipolar lobes with a higher density, $N_{outflow}$, and with the same geometry but with an outer radius, $r_{outflow}$, at which the density of the bipolar lobes dropped to the standard value. Introducing this component greatly reduced the [N~{\sc ii}], [Ar~{\sc iv}], and [K~{\sc v}], line ratios that had previously been significantly over-estimated. The final values of the two outflow parameters, $N_{outflow} = 20,000$~cm$^{-3}$ and $r_{outflow} = 2.0 \times 10^{17}$~cm (11.5$^{\prime\prime}$ at 1.17~kpc), are well constrained to $\pm$20\%, due mainly to the first two line ratios, while the third offered less of a constraint. As an example of the modeling process Table~\ref{outflow_response} lists the line ratios most responsive to the introduction of, and changes to, the outflow component.

\begin{table*}
\begin{center}
\caption{The effects of introducing and varying the parameters of the outflow component, on the most responsive density and temperature sensitive line ratios. Note how the majority of line ratios decrease as the outflow is introduced and its density increased, but then increase at very high outflow densities.}
\begin{tabular}{@{}lccccc}
\hline
Line ratio		& Observed	&  \multicolumn{4}{c}{Models with outflow density: (cm$^{-3}$)} \\
\cline{3-6}
			& &									No outflow & 10,000 & 20,000 & 40,000 \\
\hline
~[N~{\sc ii}] 5755 / (6585+6546)				& 0.0290	& 0.133	& 0.0227	& 0.0231	& 0.0278 \\
~[O~{\sc iii}] (4959+5007) / [O~{\sc ii}] (3726+3729)	& 39.1	& 401	& 259	& 87.8	& 60.6 \\
~[S~{\sc ii}] 4068 / (6731 + 6717)				& 0.485	& 2.16	& 1.13	& 0.777	& 1.38 \\
~[Ar~{\sc iv}] 4741 / 4711						& 2.01	& 7.32	& 4.96	& 4.29	& 5.10 \\
~[K~{\sc v}] 4163 / 4123						& 1.07	& 8.38	& 4.61	& 4.32	& 4.12 \\
\hline
\end{tabular}
\label{outflow_response}
\end{center}
\end{table*}

\subsection{The central star and the IR coronal lines}

Without any clear observations of the central star its parameters were particularly unconstrained at the beginning of the modeling process. While there had been many suggestions in the literature for a very high central star temperature \citep[e.g.][]{ashl88,casa00} we began the modeling process with a typical PN central star with a blackbody temperature of 90,000~K and solar abundances. While a large number of line ratios were responsive to changes in the central star temperature, the ionization structure diagnostics He~{\sc ii}~$\lambda4686$~/~He~{\sc i}~$\lambda5876$ and [O~{\sc iii}] ($\lambda$4959 + $\lambda$5007) / ($\lambda$3726 + $\lambda$3729) were particularly responsive to it and would not come close to being fit unless the blackbody effective temperature was increased to at least $\sim$150,000~K.

At this temperature, with the peak of the SED in the far-UV, further changes produced little response in any line ratios. With the exception of the total H$\beta$ luminosity of the nebula, the stellar luminosity was also very unconstrained at these temperatures. The final stellar luminosity was therefore based on the best fit to the H$\beta$ line strength. For many PNe this may be a relatively unsatisfactory result, but given the high precision to which the distance to NGC~6302 is known, its absolute H$\beta$ luminosity and therefore the luminosity of its central source, is quite well constrained.

\begin{table*}
\begin{center}
\caption{Fits to the observed infrared coronal lines from NGC~6302 using NLTE stellar model atmospheres with solar or hydrogen-deficient abundance compositions (see Section~\ref{s-centralstar}). Results for three effective temperatures are shown for each stellar abundance model (both with $\log g = 7$). Line fluxes listed as 0.0 are at least a factor of $10^4$ smaller than the observed flux. Dereddened line fluxes are taken from \citet{casa00}. All line fluxes are given in units of $10^{-12}$~erg~cm$^{-2}$~s$^{-1}$. Ionization potentials (IP) are also listed for each prior ionization stage.}
\small
\begin{tabular}{@{}lc c cc ccc cc}
\hline
Line		& IP & Observed	&	\multicolumn{7}{c}{Models} \\
\cline{4-10}
& (eV) & & \multicolumn{3}{c}{Solar composition} & & \multicolumn{3}{c}{`Typical' H-def composition} \\
 \cline{4-6} \cline{8-10}
  & & & 200kK & 220kK & 240kK & & 200kK & 220kK & 240kK \\
\hline
~[Mg {\sc v}] 5.60 $\mu$m			& 109 	& 16.9	& 6.33	& 21.0 	& 222 	& & 7.4 	& 26.6 	& 0.23 \\
~[Mg {\sc vii}] 5.51 $\mu$m		& 187 	& 13.0	& 0.67	& 1.52 	& 28.1 	& & 9.42 	& 9.58 	& 0.58\\
~[Mg {\sc viii}] 3.03 $\mu$m		& 225 	& 1.81	& 0.0 	& 0.0 	& 0.32	& & 0.91 	& 2.82 	& 4.67\\
~[Si {\sc vi}] 1.96 $\mu$m			& 167 	& 23.5	& 14.1 	& 21.7 	& 429 	& & 31.6 	& 73.4 	& 1.34\\
~[Si {\sc vii}] 2.47 $\mu$m			& 205 	& 20.5 	& 0.0 	& 0.13 	& 4.51 	& & 4.23 	& 32.1 	& 9.32\\
~[Si {\sc ix}] 3.93 $\mu$m			& 303 	& 0.030	& 0.0 	& 0.0 	& 0.0 	& & 0.002 	& 1.04 	& 139 \\
\hline
\end{tabular}
\label{coronal_line_results}
\end{center}
\end{table*}

To constrain the stellar temperature further we used observations of a group of emission lines from high-ionization stage ions (the `infrared coronal lines') that have been observed in the centre of NGC~6302 \citep[e.g.][]{ashl88,pott96,casa00}. This group includes emission from ionization stages ranging up to Si$^{8+}$ and are unique to a very small number of PNe, with NGC~6302 showing the highest ionization species ever found in a PN. These were originally suspected to originate in high-velocity shocks in the nebula \citep[e.g.][]{meab80a,lame91}, but physical arguments against such levels of shock excitation \citep{oliv96} and the lack of significantly broad wings on many of the lines \citep{casa00} argues for a photoionized origin.

The ionization potentials of many of these species lie in the extreme UV or X-ray region (see Table~\ref{coronal_line_results}) and therefore place strong restrictions on the form of the high-energy ionizing flux from the central star. As discussed above, the ionizing flux from the central star was modeled using NLTE stellar model atmosphere fluxes for a range a range of central star abundances and surface temperatures. The effects of different temperatures and abundances on the fluxes of the six highest ionization stage ions is shown in Table~\ref{coronal_line_results}. The effects of these changes on the lower ionization-stage nebula diagnostics was negligible.

Models using either a solar abundance or metal-poor stellar atmosphere were able to produce fits to many of the lines but were unable to reproduce the lines from the highest ionization stages, particularly the [Si~{\sc ix}]  3.93~$\mu$m line. The strength of this line is underestimated by a factor $\sim10^4$ even at $T_{eff} = 250,000$~K and an order-of-magnitude fit could not be obtained for any models with $T_{eff} < 400,000$~K. At such high stellar temperatures the lower ionization coronal lines (e.g. [Mg~{\sc v}] 5.60~$\mu$m and [Si~{\sc vi}] 1.96~$\mu$m) become overestimated by factors of several 100 or more. Models using hydrogen-deficient stellar atmospheres (Table~\ref{coronal_line_results}) produced substantially better fits to these lines, with the best fit obtained using a 220,000~K stellar atmosphere. The fits to the magnesium lines presented in Table~\ref{coronal_line_results} are very good. The fits to the nebular silicon lines are slightly worse, with the 2.47~$\mu$m line fitted well, but the 1.96~$\mu$m line is overestimated by a factor of three and the 3.93~$\mu$m line is overestimated by a factor of thirty.

\begin{figure}
\begin{center}
\includegraphics[width=185pt,angle=270]{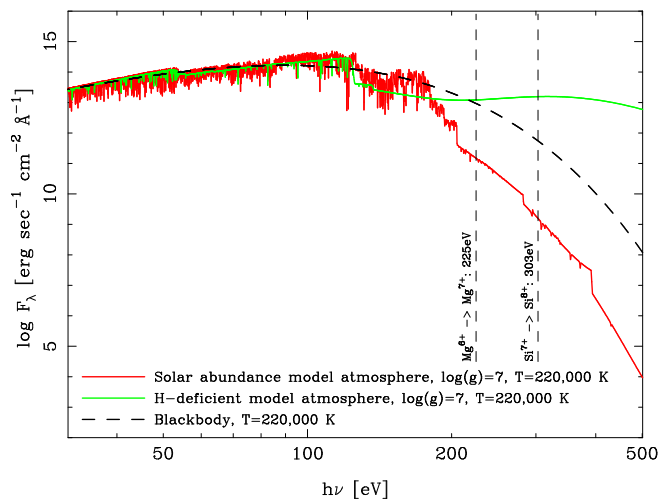}
\caption{220,000~K solar abundance and hydrogen-deficient model atmosphere SEDs compared with a blackbody at the same temperature. The frequencies corresponding to the ionization potentials of Mg$^{+6}$ and Si$^{+7}$ are shown.}
\label{compareatmospheres}
\end{center}
\end{figure}

Figure~\ref{compareatmospheres} shows the high-energy SEDs of 220,000~K solar-abundance and H-deficient model atmospheres. At the energies of the Mg$^{+6}$ and Si$^{+7}$ ionization potentials, there is a significantly higher flux in the H-deficient model atmospheres compared to the solar abundance model due to the lack of opacity from hydrogen. The absence of a number of low-abundance metals in these model atmospheres results in a reduced atmospheric opacity and therefore a higher flux at energies $> 250$~eV. Including them will likely reduce the photon flux at 330~eV and therefore resolve the problem of the over-predicted flux in the [Si~{\sc ix}] 3.93~$\mu$m line in our models.

With the exception of a blackbody \citep[which provides a better fit to the IR coronal lines, but is not an accurate representation of the ionizing flux, see e.g.][]{arms02}, the 220,000~K H-deficient stellar atmosphere provides the best fit to the observed IR coronal lines. Models with different surface gravities were tested but produced very minor changes in the model line fluxes. A value of $\log \, g = 7.0$ produced the best fit, but is poorly constrained. The fit to the higher ionization stages of the silicon ions is not ideal, but it provides a closer fit than any of the solar composition models and is strong evidence that the central star of NGC~6302 is H-deficient.

\subsection{The nebular chemistry}

Once the main nebular and central star parameters were determined by fitting the density- and temperature-sensitive line ratios, attempts were made to achieve good fits to the strengths of individual emission lines by altering the chemical abundances of the nebula, which had originally been set to literature values for NGC~6302, primarily from \citet{tsam03} and \citet{casa00}. The majority of abundances required small changes to fit the emission line fluxes, resulting in moderate variations in the nebula temperature structure which were then resolved by making other changes to the nebula structure. The sodium, magnesium, aluminium and silicon abundances were unchanged from their initial values either due to satisfactory fits, or insufficient numbers of lines to fully constrain their abundances. The final nebula abundances are listed in Table~\ref{parameters}.

A full discussion of the dust modeling process will be reserved for a future paper, but as noted earlier dust was included in these models. This caused variations in a large number of line ratios due to significant changes in the nebular ionization structure.

\section{Model results}

In this section we present the best fitting model of NGC~6302 and compare the predicted emission-line fluxes from {\sc mocassin} with those observed. The list of parameters used in the final model outlined in Section~\ref{s-model} of NGC~6302 are provided in Table~\ref{parameters}. We also note how well constrained we estimate each parameter to be. The uncertainty on each parameter is estimated as the degree to which the parameter could vary without adversely affecting the model. For nebular abundances the uncertainty is based on the differences between observed and predicted line strengths for each element.

\begin{table*}
\begin{center}
\caption{Model parameters for the best fitting final model of NGC~6302 with the origin of the value and the uncertainty on the value (the listed uncertainties do not take into account the uncertainty of $\sim$10\% on the distance to NGC~6302). All nebula abundances were constrained by the model fits, but we also note the values that remained unchanged from the initial values listed in Section~\ref{s-origabundances}. Elemental abundances are listed as [X/H], with enhancements or depletions noted in addition to the original source.}
\begin{tabular}{@{}lrll}
\hline
Model parameter		& Value	& Origin	& Uncertainty \\
\hline
Bipolar lobes:\\
Height of lobes, $H$			& $6.8 \times 10^{17}$~cm	& Estimated from images	& $\pm$10\% \\
Transition latitude, $\phi_1$	& 41.5$^\circ$				& Estimated from images	& $\pm$25\% \\
End of cone latitude, $\phi_2$	& 49.6$^\circ$				& Estimated from images	& $\pm$10\% \\
Slope of cone, $\phi_3$		& 66.0$^\circ$				& Estimated from images	& $\pm$10\% \\
Transition height, $z_p$		& $4.0 \times 10^{17}$~cm	& Estimated from images	& $\pm$25\% \\
Waist radius, $R_w$			& $1.2 \times 10^{16}$~cm	& Model fit			& $\pm$20\% \\
Lobe density				& 2000 cm$^{-3}$			& Model fit			& $\pm$50\% \\
Density profile				& constant				& Model fit			& - \\
Outflow density				& 20,000~cm$^{-3}$			& Model fit			& $\pm$20\% \\
Outflow outer radius, $r_{outflow}$ & $2.0 \times 10^{17}$~cm	& Model fit			& $\pm$20\% \\
\hline
Circumstellar disk:\\
Inner radius, $r_{in}$		& $1.2 \times 10^{16}$~cm	& Model fit			& $\pm$20\% \\
Outer radius, $r_{out}$		& $3.0 \times 10^{17}$~cm	& Not well constrained	& - \\
Length parameter, $r_d$		& $5.0 \times 10^{16}$~cm	& Model fit			& $\pm$50\% \\
Height parameter, $z_d$		& $6.0 \times 10^{15}$~cm	& Model fit			& $\pm$20\% \\
Flaring parameter, $f_d$		& 1.125					& Model fit			& $\pm$50\% \\
Characteristic density, $N_0$ 	& 80,000~cm$^{-3}$			& Model fit			& $\pm$10\% \\
\hline
Nebula abundances:\\
Hydrogen					& 1.0						& -					& - \\ 
Helium					& 0.153					& Model fit			& $\pm$10\% \\
Carbon					& $2.2 \times 10^{-4}$		& Model fit			& $\pm$25\% \\
Nitrogen					& $3.9 \times 10^{-4}$		& Model fit			& $\pm$25\% \\
Oxygen					& $5.1 \times 10^{-4}$		& Model fit			& $\pm$25\% \\
Neon					& $2.4 \times 10^{-4}$		& Model fit			& $\pm$25\% \\
Sodium					& $2.1 \times 10^{-6}$		& Model fit			& $\pm$25\% \\
Magnesium				& $1.9 \times 10^{-5}$		& Model fit			& $\pm$25\% \\
Aluminium				& $3.0 \times 10^{-8}$		& Model fit			& $\pm$50\% \\
Silicon					& $3.6 \times 10^{-5}$		& Model fit			& $\pm$50\% \\
Sulphur					& $2.5 \times 10^{-5}$		& Model fit			& $\pm$25\% \\
Chlorine					& $2.5 \times 10^{-7}$		& Model fit			& $\pm$10\% \\
Argon					& $1.2 \times 10^{-5}$		& Model fit			& $\pm$25\% \\
Potassium				& $5.0 \times 10^{-7}$		& Model fit			& $\pm$25\% \\
\hline
Central star properties:\\
Abundances				& He:C:N:O = 33:50:2:15		& Model fit			& - \\
Effective temperature		& 220,000~K				& Model fit			& $\pm$10\% \\
Luminosity				& 14,300 L$_{\odot}$		& Model fit			& $\pm$20\% \\
$\log g$					& 7.0						& Model fit			& $\pm$20\% \\
\hline
\end{tabular}
\label{parameters}
\end{center}
\end{table*}

\subsection{Fits to the emission line spectrum}
\label{s-linefits}

\begin{table*}
\begin{center}
\caption{Observed and predicted emission line fluxes for NGC~6302. The H$\beta$ line flux has been dereddened and modelled using a distance of 1.17~kpc. The UV and optical emission lines are listed in units where I(H$\beta) = 1$ while the IR lines longward of 0.9~$\mu$m are given in units of $10^{-12}$~erg~cm$^{-2}$~s$^{-1}$. See Section~\ref{s-observations} for a summary of the observations being matched here and the source of each observation. All relevant slits and apertures were simulated in {\sc mocassin} to accurately reproduce the nebular areas from which the lines were observed. References are: R94 = \citet{rowl94}, B99 = \citet{bein99}, C00 = \citet{casa00}, L01 = \citet{liu01}, G02 = \citet{grov02}, and T03 = \citet{tsam03}.}
\begin{tabular}{@{}lccc lccc}
\hline
Line						& Observed  & Model	& Reference	& Line				& Observed & Model		& Reference \\ 
\hline
~H$\beta$/$10^{-12}$~erg~cm$^{-2}$~s$^{-1}$& 776 & 852 &T03	& ~[Ne {\sc v}] 3426			& 2.33	& 6.12		& T03  \\  
~H$\beta$ 4861			& 1.00	& 1.00		& -  		& ~[Ne {\sc v}] 14.3~$\mu$m	& 634	& 1585		& B99  \\ 
~He {\sc i} 5876			& 0.175	& 0.144		& T03  	& ~[Ne {\sc v}] 24.3~$\mu$m	& 308 	& 434		& B99  \\ 
~He {\sc i} 4471			& 0.0592	& 0.0489		& T03  	& ~[Ne {\sc vi}] 7.64~$\mu$m	& 315	& 1478		& C00  \\ 
~He {\sc ii} 4686			& 0.753	& 1.18		& T03  	& ~[Na {\sc vi}] 8.64~$\mu$m	& 10.7	& 7.98		& B99   \\
~[C {\sc ii}] 157.7~$\mu$m	& 13.7	& 22.7		& L01  	& ~[Na {\sc vii}] 4.69~$\mu$m	& 4.00	& 2.13		& C00   \\ 
~[N {\sc i}] 5199			& 0.108	& 0.102		& T03    	& ~[Mg {\sc v}] 5.60 $\mu$m	& 16.9	& 26.6		& C00 \\
~[N {\sc ii}] 5755			& 0.201	& 0.205		& T03   	& ~[Mg {\sc vii}] 5.51 $\mu$m	& 13.0	& 9.58		& C00 \\
~[N {\sc ii}] 6548			& 1.74	& 2.19		& T03   	& ~[Mg {\sc viii}] 3.03 $\mu$m	& 1.81	& 2.82		& C00 \\
~[N {\sc ii}] 6584			& 5.22	& 6.67		& T03   	& ~[Al {\sc v}] 2.88~$\mu$m	& 0.0124	& 0.0742		& C00   \\ 
~[N {\sc iii}] 57.3~$\mu$m 	& 167	& 248		& L01   	& ~[Al {\sc vi}] 3.66~$\mu$m	& 0.141	& 0.102		& C00   \\  
~N {\sc iv}] 1486			& 5.34	& 5.86		& T03   	& ~[Al {\sc viii}] 3.69 $\mu$m	& 0.16	& 0.0649		& B99  \\ 
~N {\sc v} 1240				& 13.82	& 19.0		& T03   	& ~[Si {\sc ii}] 34.6 $\mu$m	& 17.4	& 325		& B99 \\
~[O {\sc i}] 5577			& 0.00232& 0.00280		& T03   	& ~[Si {\sc vi}] 1.96 $\mu$m 	& 23.5	& 73.4		& C00 \\
~[O {\sc i}] 6300			& 0.243	& 0.338		& T03   	& ~[Si {\sc vii}] 2.47 $\mu$m 	& 20.5	& 32.1		& C00 \\
~[O {\sc i}] 63.2~$\mu$m 		& 284	& 135		& L01   	& ~[Si {\sc ix}] 3.93 $\mu$m 	& 0.030 	& 1.74		& C00 \\
~[O {\sc i}] 145.5~$\mu$m 	& 10.5	& 8.27		& L01   	& ~[S {\sc ii}] 4068			& 0.145	& 0.218		& T03  \\ 
~[O {\sc ii}] 3726			& 0.300	& 0.158		& T03   	& ~[S {\sc ii}] 6716			& 0.101	& 0.0893		& T03  \\ 
~[O {\sc ii}] 3729			& 0.140	& 0.0517		& T03   	& ~[S {\sc ii}] 6731			& 0.198	& 0.191		& T03  \\
~[O {\sc ii}] 7320			& 0.0857	& 0.0866		& T03   	& ~[S {\sc iii}] 6312			& 0.0584	& 0.0328		& T03  \\
~[O {\sc ii}] 7330			& 0.0728	& 0.0713		& T03   	& ~[S {\sc iii}] 9533			& 0.717	& 0.666		& B99  \\ 
~[O {\sc iii}] 51.8~$\mu$m 	& 157	& 177		& L01   	& ~[S {\sc iii}] 18.7~$\mu$m	& 57.7	& 55.3		& B99  \\ 
~[O {\sc iii}] 88.4~$\mu$m 	& 49		& 20.2		& L01   	& ~[S {\sc iv}] 10.5~$\mu$m	& 44.0	& 18.9		& C00  \\ 
~[O {\sc iii}] 5007			& 12.9	& 13.8		& T01   	& ~[Cl {\sc iii}] 5517			& 0.00309	& 0.00391		& T03  \\ 
~[O {\sc iii}] 4959			& 4.29	& 4.62		& T03  	& ~[Cl {\sc iii}] 5537			& 0.00861& 0.0127 		& T03  \\ 
~[O {\sc iii}] 4363			& 0.396	& 0.316		& T03 	& ~[Ar {\sc ii}] 6.98~$\mu$m	& 51.0	& 45.8		& B99  \\ 
~[O {\sc iv}] 25.9~$\mu$m 	& 322	& 472		& B99  	& ~[Ar {\sc iii}] 7751			& 0.0738	& 0.0474		& G02  \\ 
~[Ne {\sc ii}] 12.8~$\mu$m	& 25.6	& 136		& C00   	& ~[Ar {\sc iii}] 7135			& 0.237	& 0.198		& T03  \\ 
~[Ne {\sc iii}] 15.6~$\mu$m	& 377	& 186		& B99  	& ~[Ar {\sc iv}] 4741			& 0.209	& 0.191		& T03  \\ 
~[Ne {\sc iii}] 36.0~$\mu$m	& 21.0	& 60.1		& B99  	& ~[Ar {\sc iv}] 4711			& 0.126	& 0.0445		& T03  \\ 
~[Ne {\sc iii}] 3967			& 0.276	& 0.338		& T03  	& ~[Ar {\sc vi}] 4.53~$\mu$m	& 29.0	& 60.3		& C00  \\ 
~[Ne {\sc iv}] 2423			& 7.95	& 4.96		& T03	& ~[K {\sc v}] 4123			& 0.00382& 0.00150		& T03 \\
~[Ne {\sc iv}] 4724			& 0.0208	& 0.0177		& T03  	& ~[K {\sc v}] 4163			& 0.00407& 0.00648		& T03 \\
~[Ne {\sc iv}] 4725			& 0.0182	& 0.0150		& T03	& ~[K {\sc vii}] 3.19~$\mu$m	& 0.279	& 0.323		& C00  \\ 
\hline
\end{tabular}
\label{line_results1}
\end{center}
\end{table*}

Predicted emission line fluxes for our best-fitting model are given in Table~\ref{line_results1}, with strengths given relative to the intrinsic dereddened H$\beta$ flux where possible. The majority of the line strengths presented are in reasonable agreement with the observations, with fits to within 30\%. In particular, nearly all the UV and optical line fluxes from \citet{tsam03} are well reproduced. We obtain a reasonable fit to the N~{\sc v}~1240~\AA\ resonance line which is often overestimated in photoionization modeling. Resonance lines must travel a much greater distance to escape from the nebula than do forbidden lines because of the large number of scatterings that they undergo and therefore can suffer greater attenuation due to dust. {\sc mocassin} treats these lines separately and their absorption by dust is accounted for in this model \citep{erco05}. 

The majority of infrared fine-structure line flux measurements are taken from spectra taken with the {\it ISO SWS} and {\it LWS}, presented by \citet{bein99} and \citet{liu01}, respectively. The remaining infrared line flux measurements were presented by \citet{casa00}, from observations made with the {\it UK Infrared Telescope} (UKIRT). We note a significantly higher discrepancy between model and observations for line fluxes measured from {\it ISO} spectra than for line fluxes measured from the {\it UKIRT} spectra. This could be due to the combined effects of a misaligned pointing used for the {\it ISO} observations and the relatively small size of the {\it ISO SWS} aperture, as in the case of the NGC~3918 modeled by \citet{erco03b}. \citet{bein99} noted a significant discontinuity between short and long wavelength sections of the SWS exposures, which they attributed to misalignment of the pointing. Combined with the 3\arcs\ pointing error on the $14 \times 20$ arcsec {\it ISO SWS} aperture positioned over the central 10\arcs~diameter peak emission region suggests errors arising from such an offset could be high. As with all our predicted line fluxes, we have reproduced the relevant aperture or slit used for the observations and therefore any inaccuracies in the observational pointing would seriously affect a comparison of line fluxes, especially for an object like NGC~6302 with such strong density and temperature contrasts in its structure. The {\it ISO LWS} aperture is much larger, making any potential pointing error smaller, and we find a smaller discrepancy between our model results than for the SWS measurements, supporting this.

Many infrared lines from neutral species are predominantly emitted from the photo-dissociation regions (PDRs) of nebulae \citep[e.g.][]{liu01,vast10} which are not treated by the current version of {\sc mocassin}. Despite this Table~\ref{line_results1} indicates that the [O~{\sc i}] 63~$\mu$m and 145~$\mu$m as well as the [C~{\sc ii}] 158~$\mu$m lines show good agreement between the predicted and observed values, despite the fact that they originate from the neutral PDR. The current version of {\sc mocassin} therefore appears to be matching these lines adequately. To complement this and for comparison with future studies we use our model to predict fluxes for two PDR lines without published fluxes, but which may be measured by future instruments. For the [C~{\sc i}] 369 and 609~$\mu$m lines we derive fluxes for the entire nebula of $7.80 \times 10^{-13}$ and $1.45 \times 10^{-13}$ ergs cm$^{-2}$ s$^{-1}$.

\subsection{Fits to the density- and temperature-sensitive line ratios}
\label{s-ratiofits}

\begin{table*}
\begin{center}
\caption{Comparison of model and observed ratios (with uncertainties) for density-, temperature-, and ionization-sensitive line ratios for NGC~6302. Also listed are the quoted densities and temperatures from the literature. References: D73 = \citet{danz73}, B82 = \citet{barr82}, R94 = \citet{rowl94}, L01 = \citet{loid01}, G02 = \citet{grov02}, T03 = \citet{tsam03}.}
\begin{tabular}{@{}lcccc}
\\
\hline
Line ratio						& Observed			& Model		& Density (cm$^{-3}$)	& Reference \\
\hline
~[C~{\sc iii}] 1907 / 1909			& $1.1 \pm 0.2$ 		& 0.522		& 16000		& B82 \\
~[O {\sc ii}] 3729 / 3726			& $0.449 \pm 0.05$		& 0.327		& 5000		& G02 \\
~[O~{\sc ii}] (3726 + 3729) / (7320 + 7330) & $2.78 \pm 0.08$	& 1.33		& 5750		& T03 \\
~[O {\sc iii}] 52 / 88 $\mu$m		& $3.20 \pm 0.12$		& 8.76		& 1380		& L01 \\
~[Ne~{\sc v}] 24.3 / 14.3 $\mu$m	& $0.49 \pm 0.05$		& 0.27		& 10000		& B99 \\
~[S {\sc ii}] 6731 / 6717			& $1.96 \pm 0.14$		& 2.14		& 12900		& T03 \\
~[Cl {\sc iii}] 5537 / 5517			& $2.54 \pm 0.28$		& 3.25		& 22450		& T03 \\
~[Ar {\sc iv}] 4741 / 4711			& $2.01 \pm 0.18$		& 4.29		& 14900		& T03 \\
~[K {\sc v}] 4163 / 4122			& $1.07 \pm 0.14$		& 4.32		& 10000		& T03 \\
\hline
Line ratio						& Observed			& Model		& Temperature (K)	& Reference \\
\hline
~[N {\sc ii}] 5755 / (6584 + 6548)	& $0.0290 \pm 0.005$	& 0.0231		& 14225	& T03 \\
~[O {\sc i}] 5577 / 6300			& $0.0126 \pm 0.0020$	& 0.0083		& 10000	& G02\\
~[O {\sc iii}] (4959 + 5007) / 4363	& $45.7 \pm 6.5$		& 58.3		& 18400	& T03 \\
~[O {\sc iii}] 4363 / 5007			& $0.029 \pm 0.010$	& 0.0229		& 17400	& D73 \\
~[S {\sc ii}] 4068 / (6731 + 6716)	& $0.485 \pm 0.082$	& 0.777		& 10000	& T03 \\
~[S {\sc iii}] 6312 / 9533			& $0.049 \pm 0.14$		& 0.0475		& 18300	& B99 \\
\hline
Line ratio						& Observed			& Model		&		& Reference \\
\hline
~He~{\sc i} 5876 / 4471			& $2.96 \pm 0.21$		& 2.94		&		& T03 \\
~He {\sc ii} 4686 / He {\sc i} 5876	& $4.30 \pm 0.30$		& 8.19		&		& T03 \\
~[O {\sc iii}] (4959 + 5007) / [O {\sc ii}] (3726 + 3729)	& $39.1\pm 6.4$ & 87.8	&		& T03 \\
\hline
\end{tabular}
\label{line_results2}
\end{center}
\end{table*}

Table~\ref{line_results2} lists the model emission line ratios that are sensitive to changes in the density, temperature, or ionization structure of the nebula, and compares them to the observed line ratios of NGC~6302. Despite the wide range of electron densities and temperatures noted in the literature for NGC~6302 we have been able to reproduce the majority of emission line ratios using our model. The fits generally show good agreement to within 20\%, with a small number only fit to within a factor of two. The most notably bad fit is the density-sensitive [K~{\sc v}] 4163 / 4122 line ratio, which is over-predicted by a factor of $\sim$4 and proved particularly difficult to fit.

The [Ar~{\sc iv}] and [K~{\sc v}] line ratios are both over-predicted by our model suggesting we have over-estimated the density of the circumstellar disk. However, many other line ratios constrained the density of the circumstellar disk and no suitable combination of parameters could be found using a lower-density circumstellar disk. Increasing the inner radius of the circumstellar disk offered a potential solution to fit these two line ratios, but increasing this parameter to $r_{in} \sim 1.5 \times 10^{16}$~cm resulted in extreme values of the [O~{\sc ii}] (3726 + 3729) / (7320 + 7330) line ratio. The introduction of an `outflow' component (see Section~\ref{s-geometry}) went a long way to resolving these discrepancies, but no simple model could be found to fit these lines accurately. Given this situation we were forced to leave these line ratios not fitted by our final model, but Section~\ref{s-inhomo} discusses a possible solution to this problem. The poor fit to the He~{\sc ii} / He~{\sc i} line ratio is also worrisome, but again Section~\ref{s-inhomo} presents a way to resolve this problem.

The majority of other line ratios are fit well and in some cases offer strong constraints on the geometry or density distribution of the nebula (see Section~\ref{s-geometry} for a discussion). 

\subsubsection{The effects of density inhomogeneities on the emission line spectrum}
\label{s-inhomo}

A number of the density- and temperature-sensitive line ratios discussed in Section~\ref{s-ratiofits}, such as the He~{\sc ii} 4686 / He~{\sc i} 5876, [Ar~{\sc iv}\ 4741 / 4711, and [K~{\sc v}] 4163 / 4122 line ratios, are not well fit by our model. Exploring the parameter space in our model (see discussion in Section~\ref{s-geometry}) revealed that almost every line ratio could be fit with some realistic combination of model parameters, but that no simple solution could be found that would fit all the line ratios. This does not necessarily suggest that a purely photoionization model solution does not exist for NGC~6302. A possible solution to this problem is that the morphology and density distribution of our model is an over-simplification of the true density distribution. The inhomogeneous appearance of the bipolar lobes of NGC~6302 (e.g. Figure~\ref{ngc6302}) suggests that the density distribution is unlikely to be smooth, with many clumps, knots and small-scale structures visible \citep[e.g.][and Figure~\ref{ngc6302}]{mats05}. Density enhancements can cause variations in density-sensitive line ratios as well as temperature-sensitive ratios though self-shielding. High densities can also induce collisional suppression of some collisionally-excited line, potentially invalidating the temperature sensitivity of some line ratios. The different densities determined from the [O~{\sc iii}] far-IR line ratio compared to the [Cl~{\sc iii}] and [Ar~{\sc iv}] line ratios, despite their similar ionization potentials (35.1eV compared to 23.8 and 40.7eV respectively), also suggests that moderate density inhomogeneities exist in the nebula and are responsible for quenching of the [O~{\sc iii}] far-IR lines by collisional de-excitation in the high-density regions \citep{liu01}.

Attempting to model this structure in detail would be impractical and largely unproductive compared to our attempts to characterise the overall structure of the nebula in Section~\ref{s-geometry}. However, exploring the influence of such inhomogeneities will allow us to test the theory that they are responsible for the discrepancies between our model and the observations. To simulate these density inhomogeneities a number of `knots' of various sizes and densities were simulated in our model using {\sc mocassin}'s subgrid feature to resolve small-scale structures. For simplicity these knots were simulated as spherical over-densities in the structure of either the bipolar lobes or the circumstellar disk. Two parameters were used to model these structures, $f_\rho$, a density factor representing the peak over-density of the knot, and $r_{knot}$, the radius of the knot. All the knots were modeled using a $r^{-1}$ radial density profile. These parameters were allowed to vary over the ranges $1.0 < f_\rho < 3.0$ and $10^{16} < r_{knot} / \mathrm{cm} < 10^{17}$, the latter of which was estimated from images of the nebula. The knots were distributed randomly across either the bipolar lobes, the circumstellar disk, or both, and the total number of knots was varied from 5--50.

The effect of these density enhancements on the nebula ionization structure depends on their position in the nebula and their density, but their main effect is to create shielded regions where lower ionization stages can exist. The most notable effects are therefore on the line ratios that probe the ionization structure of the nebula, particularly the He~{\sc ii}~/~He~{\sc i} line ratio which is significantly reduced if knots are introduced into the circumstellar disk. The other ionization-dependent line ratio, [O~{\sc iii}] (4959 + 5007) / [O~{\sc ii}] (3726 + 3729), was also found to decrease if knots were added to either the lobes or the disk, which had the effect of worsening the fit to this line ratio. However the magnitude of this change was smaller than that of the helium line ratio, suggesting it could be countered with other model changes.

Changes to density-sensitive line ratios were also observed when knots were added to the bipolar lobes, particularly the [Cl~{\sc iii}], [Ar~{\sc iv}], and [K~{\sc v}] line ratios that all probe high density regions, which were found to move closer to the observed ratios. The [O~{\sc ii}] and [S~{\sc ii}] line ratios, sensitive to lower-density regions, were less responsive. Adding knots to the circumstellar disk did not induce significant changes in these density-sensitive line ratios, most probably because either the upper critical densities of these line ratios are lower than the typical densities in the circumstellar disk or because the majority of the circumstellar disk is in a low ionization state.

In summary, introducing small-scale density inhomogeneities or `knots' in the nebula structure could resolve some of the remaining discrepancies between our model and the observations. Knots in the circumstellar disk could resolve the over-predicted ionization-sensitive helium line ratio, while knots in the bipolar lobes could resolve differences in line ratios sensitive to high-densities. Despite this, a number of the remaining line ratios could not be matched within the parameter space explored (e.g. the low-density-sensitive [O~{\sc ii}] (3726 + 3729) / (7320 + 7330) line ratio).

\subsection{The nebular ionization structure}

\begin{table}
\begin{center}
\caption{Integrated ionic abundance ratios for He, C, N, O, Ne, S, Cl, and Ar, derived from model ionic fractions and compared to the results of \citet{tsam03}.}
\begin{tabular}{@{}lcc}
\hline
Ionic ratio				& Model		& Observed\\
\hline
He$^+$ / H$^+$		& 0.0645		& 0.0658\\
He$^{2+}$ / H$^+$		& 0.0477		& 0.0696\\
C$^{2+}$ / H$^+$		& 1.82(-5)		& 3.08(-5)\\
C$^{3+}$ / H$^+$		& 9.51(-5)		& 1.43(-5)\\
N$^+$ / H$^+$			& 1.34(-4)		& 4.48(-5)\\
N$^{2+}$ / H$^+$		& 7.44(-5)		& 1.03(-4)\\
N$^{3+}$ / H$^+$		& 3.75(-5)		& 7.80(-5)\\
N$^{4+}$ / H$^+$		& 4.06(-5)		& 9.45(-5)\\
O$^+$ / H$^+$			& 3.00(-4)		& 1.67(-4)\\
O$^{2+}$ / H$^+$		& 4.17(-5)		& 9.35(-5)\\
O$^{3+}$ / H$^+$		& 1.88(-5)		& 8.22(-5)\\
Ne$^{2+}$ / H$^+$		& 1.19(-5)		& 2.65(-5)\\
Ne$^{3+}$ / H$^+$		& 5.04(-5)		& 3.53(-5)\\
S$^+$ / H$^+$			& 1.28(-6)		& 8.20(-7)\\
S$^{2+}$ / H$^+$		& 1.98(-6)		& 1.57(-6)\\
Cl$^{2+}$ / H$^+$		& 1.13(-7)		& 2.76(-8)\\
Ar$^{2+}$ / H$^+$		& 7.10(-7)		& 6.51(-7)\\
Ar$^{3+}$ / H$^+$		& 3.39(-7)		& 7.75(-7)\\
\hline
\end{tabular}
\label{ion_compare}
\end{center}
\end{table}

A comparison of the relative ionic abundances in our model with those determined by \citet{tsam03} is shown in Table~\ref{ion_compare}. The ionization structure derived by those authors is in good agreement with our model, though there are a number of significant differences. The ionic fractions of O$^{3+}$ differ by a factor of four, while the fraction of Cl$^{2+}$ in our models is a factor of five larger than the empirical observational value. Fits to the [O~{\sc iv}] and [Cl~{\sc iii}] lines in Table~\ref{line_results1} are generally good and don't show major discrepancies. The likely explanation for this difference is that the model ionic abundances reported here are for the entire nebula, while those reported by \citet{tsam03} only apply to the parts of the nebula caught under the slit.

\begin{table*}
\begin{center}
\caption{Nebular-averaged fractional ionic abundances for the NGC 6302 model. For each element the upper row is for the optically thick circumstellar disk and the lower row is for the optically thin bipolar lobes.}
\footnotesize 
\begin{tabular}{@{}lllllllll}
\\ 
\hline 
 & \multicolumn{8}{c}{Ion} \\ 
\cline{2-9} 
Element & \multicolumn{1}{c}{{\sc i}} & \multicolumn{1}{c}{{\sc ii}} & \multicolumn{1}{c}{{\sc iii}} & \multicolumn{1}{c}{{\sc iv}} & \multicolumn{1}{c}{{\sc v}} & \multicolumn{1}{c}{{\sc vi}} & \multicolumn{1}{c}{{\sc vii}} & \multicolumn{1}{c}{{\sc viii}} \\ 
\hline 
H & 0.991 & 0.937(-2) \\ 
 & 0.272 & 0.728 \\ 
He & 0.996 & 0.369(-2) & 0.236(-3) \\ 
 & 0.142 & 0.506 & 0.352 \\ 
C & 0.983 & 0.150(-1) & 0.157(-2) & 0.821(-4) & 0.408(-4) & 0.255(-6) & 0.504(-9) \\ 
 & 0.366(-2) & 0.497 & 0.255 & 0.910(-1) & 0.151 & 0.281(-2) & 0.346(-4) \\ 
N & 0.982 & 0.179(-1) & 0.994(-4) & 0.502(-4) & 0.543(-4) & 0.243(-4) & 0.177(-7) & 0.002(-9) \\ 
 & 0.137 & 0.479 & 0.135 & 0.495(-1) & 0.710(-1) & 0.129 & 0.280(-3) & 0.269(-6) \\ 
O & 0.981 & 0.187(-1) & 0.111(-3) & 0.501(-4) & 0.588(-4) & 0.199(-4) & 0.235(-5) & 0.159(-9) \\ 
 & 0.311 & 0.316 & 0.136 & 0.391(-1) & 0.737(-1) & 0.935(-1) & 0.317(-1) & 0.468(-5) \\ 
Ne & 0.983 & 0.153(-1) & 0.105(-2) & 0.443(-4) & 0.803(-4) & 0.996(-5) & 0.260(-6) & 0.229(-7) \\ 
 & 0.196(-1) & 0.210 & 0.528 & 0.365(-1) & 0.134 & 0.671(-1) & 0.422(-2) & 0.851(-3) \\ 
Na & 0.883 & 0.116 & 0.133(-2) & 0.583(-4) & 0.727(-4) & 0.196(-4) & 0.399(-5) & 0.732(-6) \\ 
 & 0.315(-1) & 0.260 & 0.450 & 0.473(-1) & 0.878(-1) & 0.681(-1) & 0.411(-1) & 0.114(-1) \\ 
Mg & 0.736 & 0.264 & 0.284(-3) & 0.213(-4) & 0.663(-4) & 0.410(-4) & 0.753(-5) & 0.119(-5) \\ 
 & 0.271(-1) & 0.586 & 0.140 & 0.205(-1) & 0.635(-1) & 0.904(-1) & 0.529(-1) & 0.167(-1) \\ 
Al & 0.247 & 0.752 & 0.512(-3) & 0.190(-4) & 0.653(-4) & 0.435(-4) & 0.107(-4) & 0.171(-5) \\ 
 & 0.162(-2) & 0.612 & 0.123 & 0.334(-1) & 0.601(-1) & 0.806(-1) & 0.623(-1) & 0.220(-1) \\ 
Si & 0.249 & 0.750 & 0.403(-3) & 0.421(-4) & 0.555(-4) & 0.195(-4) & 0.653(-5) & 0.172(-5) \\ 
 & 0.134(-2) & 0.734 & 0.215(-1) & 0.458(-1) & 0.510(-1) & 0.600(-1) & 0.561(-1) & 0.248(-1) \\ 
S & 0.690 & 0.305 & 0.472(-2) & 0.140(-4) & 0.427(-4) & 0.561(-4) & 0.237(-4) & 0.364(-5) \\ 
 & 0.597(-3) & 0.390 & 0.341 & 0.324(-1) & 0.331(-1) & 0.609(-1) & 0.966(-1) & 0.373(-1) \\ 
Cl & 0.911 & 0.862(-1) & 0.255(-2) & 0.722(-4) & 0.395(-4) & 0.292(-4) & 0.195(-4) & 0.432(-5) \\ 
 & 0.221(-2) & 0.447 & 0.244 & 0.931(-1) & 0.262(-1) & 0.359(-1) & 0.857(-1) & 0.558(-1) \\ 
Ar & 0.967 & 0.325(-1) & 0.722(-3) & 0.344(-4) & 0.101(-4) & 0.514(-4) & 0.687(-4) & 0.591(-5) \\ 
 & 0.309(-1) & 0.198 & 0.432 & 0.973(-1) & 0.121(-1) & 0.405(-1) & 0.135 & 0.464(-1) \\ 
K & 0.874 & 0.125 & 0.654(-3) & 0.618(-4) & 0.271(-4) & 0.402(-4) & 0.419(-4) & 0.782(-5) \\ 
 & 0.482(-1) & 0.387 & 0.265 & 0.750(-1) & 0.173(-1) & 0.309(-1) & 0.920(-1) & 0.683(-1) \\ 
\hline 
\end{tabular}
\label{ionic_abundances}
\end{center}
\end{table*}
\normalsize 

\begin{figure*}
\begin{center}
\includegraphics[height=500pt,angle=270]{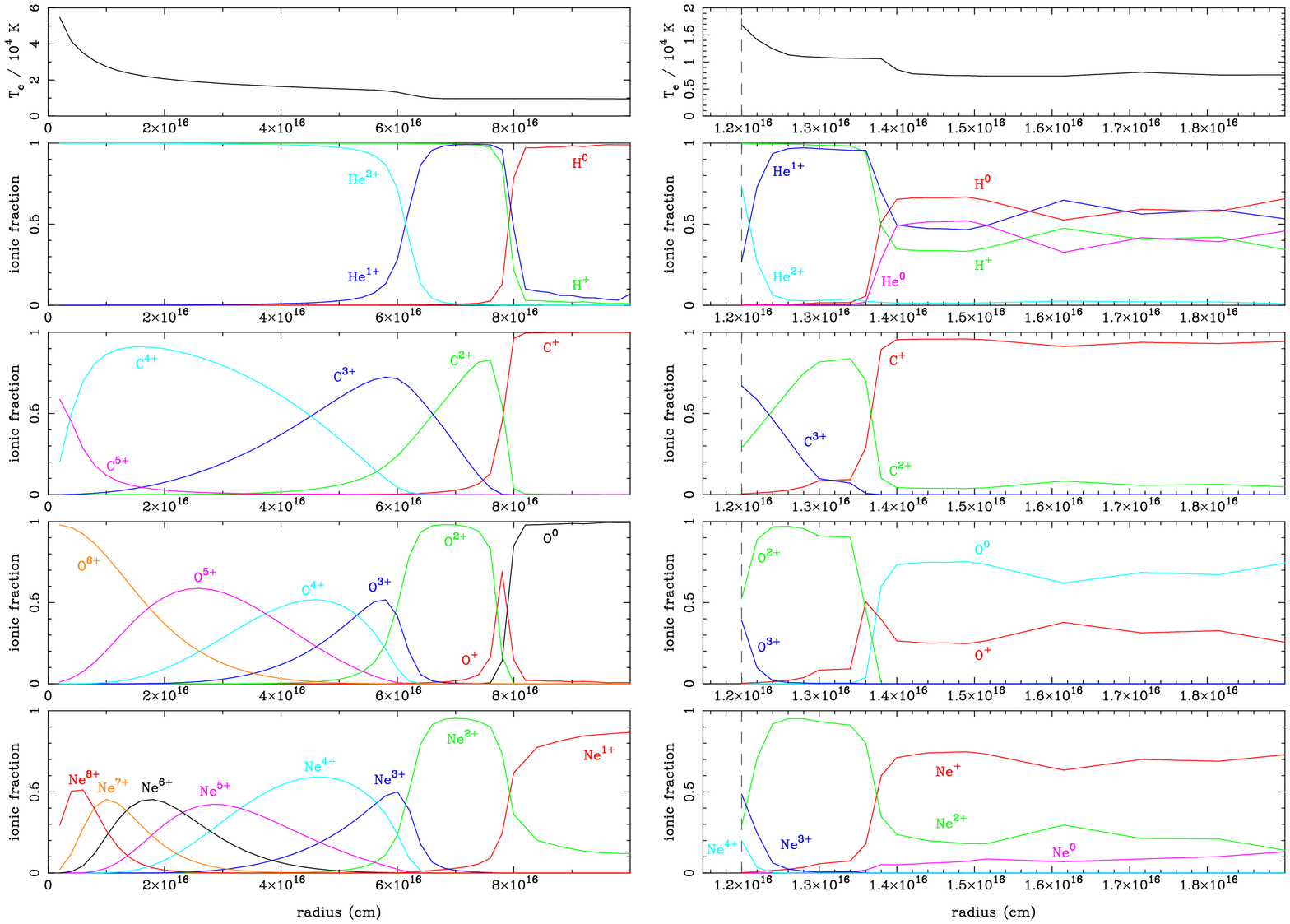}
\caption{Ionic fractions and electron temperatures for two sightlines through the model nebula. {\it Left}: along the $z$-axis through the bipolar lobes; {\it right}: along the $x$- or $y$-axis through the circumstellar disk. Note the different spatial dimensions on each. Ionization fractions are shown for helium and hydrogen (2nd panel from top), carbon (middle panel), oxygen (2nd panel from bottom), and neon (bottom panel). The dashed line in the right-hand panels indicates the edge of the circumstellar disk.}
\label{ionicfractions}
\label{temp_profiles}
\end{center}
\end{figure*}

Table~\ref{ionic_abundances} provides complete fractional ionic abundances for our NGC~6302 model for both the circumstellar disk and bipolar lobes. Figure~\ref{ionicfractions} shows the ionic fractions for a number of elements along two sightlines that pass through the bipolar lobes (along the $z$-axis) and the circumstellar disk (along the $x$ and $y$ axes). In the bipolar lobes nearly all ionization stages are seen somewhere, with a clear ionization sequence from the inner heavily ionized regions to the outer parts of the lobes where some material is neutral. Hydrogen is 73\% ionized over the entire lobes, while helium is 51\% singly ionized and 35\% doubly ionized. In the circumstellar disk the ionization front occupies a thin region at the inner edge of the disk, almost one hundred times closer to the central star than in the bipolar lobes. Because of this the vast majority of the circumstellar disk is neutral with hydrogen $>$99\% neutral in the disk. This is in agreement with the H$\alpha$ images of \citet{mats05} that show very little emission from the disk of NGC~6302. Comparison between the [Ne~{\sc v}] images presented by \citet{bohi98} and our model Ne$^{4+}$ fractions also shows good agreement, with 13\% of the neon in the bipolar lobes in the form of Ne$^{4+}$, but less than 0.1\% in the circumstellar disk.

Beyond the $H^+$ ionization front at a depth of $\sim 1.4 \times 10^{16}$~cm the disk is optically thick and our model had difficulties converging due to the lack of photons. This is evident from the noise in the ionic fractions, predominantly of neutral species, beyond the ionization front. However, we found very little variation between Monte Carlo simulations in the total neutral and singly-ionized fractions of the majority of species, indicating that this noise did not affect our predicted emission line fluxes.

\subsection{The nebular temperature structure}
\label{s-temp_structure}

Mean electron temperatures weighted by ionic abundance have been calculated and are given in Table~\ref{ionic_temperatures}. The ionic temperatures for the same ions in different regions of the model show the influence of the density on the shielding of ions and therefore the temperature structure. In the optically thick circumstellar disk the neutral species all have temperatures 1000--3000~K, while those species in the optically thin bipolar lobes have temperatures around 7,000--10,000~K. The ionic temperatures increase towards higher ionization stages, but increase faster in the disk than in the lobes.

Because {\sc mocassin} does not currently treat the neutral region fully, the temperatures of the neutral species likely represents the narrow transition region between ionized and neutral regions and not the highly optically thick and neutral region deep in the circumstellar disk. Beyond the ionization front, temperatures in the circumstellar disk drop to well below 1000~K and some cells have temperatures $< 100$~K. In these conditions complex molecules are likely to form, as observed by \citet{pere07}, but we are unable to model this with the current version of {\sc mocassin}.

Electron temperature profiles are shown in Figure~\ref{temp_profiles} as a function of radius along two axes. In the inner parts of the lobes the temperature reaches in excess of $5 \times 10^4$~K (due to the lack of neutral H$^0$ which would otherwise cause Ly-$\alpha$ cooling) but drops to 10,000~K by a radial distance of $\sim 7 \times 10^{16}$~cm. The temperatures in the disk do not reach such high levels, but do show the characteristic drops at the He$^{2+}$, He$^+$, and H$^+$ ionization fronts (the latter pair of which, at a radius of $\sim 1.4 \times 10^{16}$~cm, have coincident ionization fronts as would be expected for such a hot central star). This range of temperatures is in full agreement with that found by many previous authors (e.g. Table~\ref{line_results2}), namely a nebula with a mean ion-weighted electron temperature of 19,400~K in the bipolar lobes, 3,800~K in the circumstellar disk and 12,000~K averaged over the entire nebula.

\begin{table*}
\begin{center}
\caption{Mean electron temperatures (K) weighted by ionic species for the NGC 6302 model. For each element the upper row is for the optically thick circumstellar disk and the lower row is for the optically thin bipolar lobes.}
\footnotesize 
\begin{tabular}{@{}lrrrrrrrr}
\\ 
\hline 
 & \multicolumn{8}{c}{Ion} \\ 
\cline{2-9} 
Element & \multicolumn{1}{c}{{\sc i}} & \multicolumn{1}{c}{{\sc ii}} & \multicolumn{1}{c}{{\sc iii}} & \multicolumn{1}{c}{{\sc iv}} & \multicolumn{1}{c}{{\sc v}} & \multicolumn{1}{c}{{\sc vi}} & \multicolumn{1}{c}{{\sc vii}} & \multicolumn{1}{c}{{\sc viii}} \\ 
\hline 
H & 2018 & 7367 \\ 
 & 7742 & 12682 \\ 
He & 2237 & 8094 & 14838 \\ 
 & 7468 & 8874 & 15415 \\ 
C & 896 & 4736 & 5754 & 15740 & 16164 & 17461 & 20243 \\ 
 & 8088 & 8164 & 9769 & 15372 & 18542 & 22326 & 36914 \\ 
N & 1177 & 4817 & 11435 & 15701 & 15813 & 16424 & 18225 & 20780 \\ 
 & 7259 & 8478 & 10270 & 13823 & 16863 & 18911 & 24343 & 43851 \\ 
O & 1428 & 4596 & 10912 & 15707 & 15807 & 16330 & 18343 & 20485 \\ 
 & 7681 & 8764 & 10358 & 14707 & 16787 & 18489 & 20652 & 29473 \\ 
Ne & 944 & 4421 & 8724 & 15628 & 15861 & 16720 & 19346 & 21347 \\ 
 & 6964 & 7740 & 9147 & 14256 & 17270 & 19222 & 22298 & 28040 \\ 
Na & 2041 & 4061 & 8681 & 15474 & 15785 & 16139 & 17509 & 19813 \\ 
 & 7804 & 7818 & 9164 & 13299 & 16492 & 18159 & 19417 & 21362 \\ 
Mg & 2002 & 3720 & 7517 & 15648 & 15750 & 15945 & 16777 & 18939 \\ 
 & 8602 & 8358 & 9933 & 13272 & 15950 & 17627 & 19022 & 20724 \\ 
Al & 2785 & 2621 & 6110 & 15370 & 15742 & 15893 & 16545 & 18455 \\ 
 & 8533 & 8452 & 9680 & 11991 & 15757 & 17382 & 18807 & 20286 \\ 
Si & 2806 & 2724 & 8260 & 15705 & 15753 & 16116 & 17115 & 18994 \\ 
 & 8720 & 8718 & 11983 & 13996 & 16281 & 17937 & 18963 & 20207 \\ 
S & 440 & 3389 & 4853 & 15398 & 15715 & 15781 & 16219 & 17532 \\ 
 & 7954 & 7967 & 9250 & 11327 & 14722 & 16500 & 18216 & 19562 \\ 
Cl & 1112 & 4072 & 4904 & 14724 & 15693 & 15801 & 16278 & 18064 \\ 
 & 7986 & 8057 & 9390 & 11455 & 15087 & 16620 & 18059 & 19445 \\ 
Ar & 848 & 4356 & 8356 & 10308 & 15681 & 15732 & 15911 & 16941 \\ 
 & 7037 & 7624 & 8789 & 10223 & 13535 & 15601 & 17615 & 19354 \\ 
K & 2113 & 4509 & 9048 & 13819 & 15641 & 15738 & 15946 & 17080 \\ 
 & 8079 & 8046 & 9385 & 10977 & 14358 & 15994 & 17553 & 19050 \\ 
\hline 
\end{tabular}
\label{ionic_temperatures}
\end{center}
\end{table*}
\normalsize 

\subsection{Comparison with observations of the central star}

\begin{table}
\begin{center}
\caption{Predicted unreddened, reddened, and observed photometry of the central star of NGC~6302. All magnitudes are presented on the Vega system. Reddened magnitudes were calculated using a \citet{howa83} extinction law with $R=3.1$ and $A_V = 8.22$.}
\begin{tabular}{@{}lccccc}
\hline
Filter		& $\lambda$ 	& FWHM		& Observed		& \multicolumn{2}{c}{Model} \\
\cline{5-6}
		& (\AA)		& (\AA)		& 				& $m_0$ & $m$ \\
\hline
F469N	& 4688.1		& 37.2		& $22.63 \pm 0.10$	& 12.57		& 22.60 \\
F673N	& 6765.9		& 100.5		& $20.08 \pm 0.07$	& 13.73		& 20.05 \\
\hline
\end{tabular}
\label{photometry}
\end{center}
\end{table}

The central star of NGC~6302 was first detected in HST observations by \citet[][and Figure~\ref{ngc6302}]{szys09}, who attempted photometry in a number of narrow filters. Due to nebular contamination in many of the filters only two of these provide a reliable measurement of the stellar magnitude. Table~\ref{photometry} provides a comparison of their measured values with predictions from our model. To calculate the unreddened stellar magnitudes the $T_{eff} = 220,000$~K, $L_\star = 14,300 L_\odot$ model atmosphere was convolved with the relevant HST filter profiles and a spectrum of Vega. The resulting magnitudes are $\sim$6-10 magnitudes brighter than the observed photometry suggesting a considerable amount of extinction \citep[e.g.][]{mats05}. Using an $R=3.1$ extinction law in the form given by \citet{howa83} we reddened the model photometry, varying the extinction to obtain agreement with the observed magnitudes. A very good fit was found using $A_V = 8.22$, fitting both magnitudes to within the observational uncertainties. This extinction is slightly larger than the $A_V \sim 6.7$ found by \citet{szys09}, but in agreement with other literature values \citep[e.g.][estimated $A_{H\alpha} = 5-7$, approximately $A_V = 6-8$]{mats05}. The good agreement between observed and predicted photometric magnitudes and colours is strong verification of the central star luminosity and effective temperature we derived.

As illustrated by Figure~\ref{compareatmospheres} the high temperature and luminosity of the central star combined with its hydrogen-deficient nature results in a not-insignificant flux at UV and X-ray energies such that the central star of NGC~6302 should be a detectable X-ray source. This provides an observable test of both the high temperature and hydrogen deficiency of the central star. We have calculated the luminosity of the central star in the {\it Chandra} (0.5--8.0 keV) and {\it XMM-Newton} (0.3--4.5 keV) bands as $5 \times 10^{34}$~erg~s$^{-1}$ and $6 \times 10^{35}$~erg~s$^{-1}$, respectively. X-ray absorption from neutral hydrogen in the circumstellar disk is calculated using the extinction of $A_V = 8.22$ estimated above, and the conversion to X-ray absorbing cross-section from \citet{ryte96}. At a distance of 1.17~kpc this predicts absorbed fluxes of $\sim 5 \times 10^{-14}$~erg~s$^{-1}$~cm$^{-2}$ for both X-ray bands (absorption due to neutral hydrogen is stronger at lower energies, off-setting the lower energy range of the {\it XMM-Newton} band for this soft X-ray source). Unfortunately the uncertainty on the predicted flux is relatively high due to uncertainties on the exact form of the stellar spectrum and the column of absorbing gas along the line of sight. However a H-rich central star of an equivalent effective temperature would be a significantly weaker X-ray source such that the detection of an X-ray point source in NGC~6302 would be sufficient to confirm the H-deficient nature of the central star. Since X-rays may also originate from shocked gas or collimated flows in the central cavity of the nebula \citep[e.g.][]{kast08,park09}, such observations must also distinguish between diffuse and point-source X-ray emission if the X-ray detection of the central star is to be observationally tested.

\section{Discussion}
\label{discuss-model}

In this section we discuss the properties of the nebula and central star modeled here with particular focus on the parts of the model that are most strongly constrained by the observations. We discuss these in the context of previous studies of NGC~6302 and the general properties of planetary nebulae. Finally we suggest an evolutionary scenario for the formation and history of NGC~6302.

\subsection{The nebula structure}
\label{discuss-structure}

NGC~6302 has long been known as a highly inhomogeneous nebula and our model matches this. Between the bipolar lobes and the inner edge of the circumstellar disk we model a density contrast of two orders of magnitude, from 2000~cm$^{-3}$ in the lobes to 300,000~cm$^{-3}$ in the circumstellar disk. This was necessary to reproduce many of the diagnostic emission line ratios, which were impossible to fit without including both the low and high density parts of the model. Such high densities are rare in PNe \citep[e.g.][]{oste06}. In their study of PN densities \citet{tsam03}, NGC~6302 consistently displayed the highest densities from a range of diagnostic line ratios, with only NGC~5315 and IC~4191 showing comparably high densities from their sample.

\begin{table}
\begin{center}
\caption{Model properties of the best fitting final model of NGC~6302.}
\begin{tabular}{@{}lrl}
\hline
Model property				& Value						& Uncertainty \\
\hline
Volume of bipolar lobes		& $9.30 \times 10^{53}$~cm$^{3}$	& $\pm$20\% \\
Mass of bipolar lobes		& 2.47~M$_{\odot}$				& $\pm$30\% \\
Volume of circumstellar disk	& $1.40 \times 10^{53}$~cm$^{3}$	& $\pm$50\% \\
Mass of circumstellar disk		& 2.22~M$_{\odot}$				& $\pm$50\% \\
\hline
Total modeled nebular mass	& 4.69~M$_{\odot}$				& $\pm$40\% \\
Total ionized mass			& 1.82~M$_{\odot}$				& $\pm$20\% \\
\hline
\end{tabular}
\label{properties}
\end{center}
\end{table}

Derived total volumes and masses for our model nebula are listed in Table~\ref{properties}. The total ionized mass of our model is $\sim$1.82~M$_{\odot}$ (predominantly located in the bipolar lobes), much higher than the typical ionized masses of Type~I PNe \citep[e.g. 0.54~M$_{\odot}$ for six Magellanic Cloud Type~I PNe,][]{barl87}. The total mass of nebular material modeled (not including dust) is $\sim$4.7~M$_{\odot}$, also notably higher than the typical masses of PNe \citep[][studied 96 PNe and found masses in the range $0.01 < \mathrm{M}_i / \mathrm{M}_{\odot} < 0.8$]{pott96b}, but not implausible. The final disk radius value of $3 \times 10^{17}$~cm was constrained from fitting the infrared SED (not discussed here), but is unconstrained from photoionization modeling. The total disk mass is in good agreement with previous observations. \citet{pere07} estimate the mass of the molecular region of NGC~6302 to be as much as $1.8 \pm 0.9$~M$_{\odot}$, in good agreement with our disk mass of $\sim$2.2~M$_{\odot}$, of which 97\% (2.1~$M_{\odot}$) is neutral. However, \citet{dinh08} derive a significantly lower molecular mass of 0.12~M$_{\odot}$ (scaled to a distance of 1.17~kpc), but note that this may be a lower limit if the CO temperature in the disk is higher than the 30~K they assume.

During the modeling process we attempted to model the circumstellar disk as a torus but found that the height of the circumstellar material was well constrained by the observations and that a flattened circumstellar disk was a better fit. Evidence for an extended disk-like morphology comes from \citet{bohi94} who observed differential extinction in the western lobe, which he suggested is viewed through the outer parts of the disk. \citet{icke03} suggested that a warped disk structure is also necessary to form the multiple lobes seen in NGC~6302. A thin gas disk and a molecular torus could both be present in NGC~6302 if the torus of dust and molecular material were embedded within the outer regions of a sufficiently flared circumstellar disk. It is possible that any once torus-like structure would become thin and disk-like due to radiation pressure and stellar winds.

\subsection{Nebular abundances}

We fit our model of NGC~6302 with a helium abundance of 0.153, which is high even for a Type~I PN \citep{king94}, indicative of a large amount of stellar processing in the progenitor star. According to the dredge-up models of \citet{beck80}, nebular material with this He abundance should correspond to a star with an initial mass $> 4$~M$_{\odot}$. The O/H abundance we derive for NGC~6302 ($5.1 \pm 1.3 \times 10^{-4}$) agrees within the uncertainty limits with the mean O/H ratio for Galactic disk PNe by \citet{king94} and with the solar O/H ratios of \citet{caff08} and \citet{aspl09}. We do not confirm the depletion of oxygen by a factor of two that was found by \citet{pott99} and \citet{tsam04} from empirical abundance analyses of collisionally excited UV, optical, and infrared lines. We attribute this difference to the neglect by empirical analyses of the large fraction of oxygen that resides in very high ionization stages in this unusually highly ionized nebula (see Table~\ref{ionic_abundances}).

We find a C/O ratio of 0.43 for NGC~6302, indicating that the nebula is predominantly O-rich, in agreement with all other measurements of the C/O ratio in the literature (e.g. C/O$ = 0.20$, \citealt{alle81}, C/O$ = 0.26$, \citealt{pott99}, C/O$ = 0.88$, \citealt{casa00}, and C/O$ = 0.31$, \citealt{tsam03}), with the majority of values $< 0.5$. Our derived C/H ratio of $2.2 \times 10^{-4}$ is 80\% of the solar value listed by \citet{aspl09}, indicating only a slight depletion of carbon, if any. However, the N/H ratio we derive for NGC~6302 ($3.9 \times 10^{-4}$) is a factor of 5.7 larger than the solar value given by \citet{aspl09}, too large to have been produced by secondary conversion of initial carbon, while the combined (C+N+O)/H ratio for NGC~6302 of $11.2 \times 10^{-4}$ is 35\% larger than the solar value of $8.28 \times 10^{-4}$. We conclude that primary enrichment of nitrogen occurred in the precursor star of NGC~6302, via 3rd dredge-up enhancement of carbon, followed by hot bottom burning CN-cycle conversion of dredged-up carbon to nitrogen.

Several elements appear particularly depleted, such as magnesium (by a factor of two compared to the solar value) and aluminium (by a factor of one hundred). This is strong evidence for depletion onto dust grains, and given the evidence for large amounts of dust in NGC~6302 \citep[e.g.][]{mols01}, is not unexpected.

\subsection{The central star}

The observation of emission lines from very high ionization stages and the nebula's Type~{\sc i} status has long suggested that the central star of NGC~6302 was extremely hot. The hottest PN central stars directly measured to date have surface temperature of up to 180,000~K \citep{wern97}, which though high, are lower than all estimated temperatures for NGC~6302's central star \citep[e.g.][estimated a temperature of 430,000~K]{ashl88}.

Using a H-deficient stellar model atmosphere we find a best fit to the observed emission-line spectrum with central star properties $T_{eff}$~=~220,000~K, $\log g = 7.0$, and L$_{\star} =$~14,300~L$_{\odot}$. This is the lowest temperature ever estimated for the central star of NGC~6302, due to our use of a H-deficient stellar atmosphere which has a larger flux at the energies necessary to significantly populate the highest ionization states from which emission lines are observed (see Figure~\ref{compareatmospheres}). However, this luminosity is significantly larger than previous estimates, or our estimate of the total luminosity of 5700~L$_\odot$ for NGC~6302, in Section~\ref{s-starlum}, from integrating its dereddened observed emission across all wavelengths longwards of 1200~\AA. That this is only 40\% of the stellar luminosity needed by our bipolar model to match the observed line and continuum fluxes suggests that 60\% of the stellar luminosity escapes through optically thin regions of the nebula.

The observed expansion of NGC~6302's bipolar lobes \citep{meab08,szys11} suggests it once had a fast wind that could affect the stellar spectrum. However strong mass-loss is usually only seen at lower effective temperatures and lower surface gravities, e.g. $\log g = 3.0 - 5.5$ \citep[e.g.][find no evidence for stellar winds in the UV spectra of very hot PN central stars that \citealt{bene09} found to have surface gravities of $\log g = 6.9 - 7.3$]{bohl82}. As we will show in Section~\ref{s-tracks}, evolutionary tracks for LTP and VLTP models imply a surface gravity of $\log g \sim 6.6$ for the $T_{eff}$ and $L_\star$ we have derived, and this can also be derived from simple stellar relations by assuming a typical central star mass. Furthermore \citet{casa00} could find no evidence for a hot wind or a wind blown cavity from observations of the IR coronal lines originating in the center of NGC~6302.

If the central star of NGC~6302 is H-deficient this implies that it has undergone some sort of late thermal pulse during its evolution, which has consumed the majority of the remaining surface hydrogen \citep[e.g.][]{herw99}. There are three possible scenarios for this: the `after final thermal pulse' (AFTP) scenario, where an extra thermal pulse is experienced at the top of the AGB \citep{herw01}; the `late thermal pulse' (LTP) that occurs during the post-AGB evolution while H-burning is still active \citep{bloc01}; or the `very late thermal pulse' (VLTP) experienced by a hot white dwarf during its early cooling phase \citep{iben83}. Only the VLTP scenario consumes the majority of the remaining hydrogen in the central star, while the other two scenarios suggest a dilution of the surface hydrogen to very low levels. We are currently unable to compare model atmospheres with H-deficient abundances with H-poor abundances, so we cannot use this method to separate these scenarios.

H-deficient central stars include the Wolf-Rayet central stars \citep[$\lbrack$WC$\rbrack$, e.g.][]{hama97} and the hotter and rarer PG~1159 stars \citep{wern06}. [WC]-type stars have been suggested to be the evolutionary precursors of PG~1159 stars \citep{gorn00}, with the earlier stages characterised by strong stellar winds and interaction with the remaining nebulous material, and the latter stage exhibiting an often isolated central star at its hottest temperature. The high temperature and high surface gravity of the central star of NGC~6302 suggests it is similar to a PG~1159 star, and the presence of the remaining nebulous material can be explained if the star were particularly massive and has therefore evolved to this temperature before the nebula has dispersed.

There are many similarities between NGC~6302 and other PNe with H-deficient central stars that support this assertion. \citet{zijl01a} notes that nearly all PNe with [WC]-type central stars feature both O-rich and C-rich circumstellar material (i.e. dual-dust chemistry), as does NGC~6302 \citep{mols01,cohe01}. However, most H-deficient nebulae exhibit much stronger PAH (polycyclic aromatic hydrocarbon) emission than NGC~6302, though this could be due to the very low C/O ratio in NGC~6302 \citep[e.g.][show that the production of PAHs is greatly reduced at low C/O ratios]{guzm11}. Another similarity between NGC~6302 and PNe with H-poor or H-deficient central stars is that they both show highly inhomogeneous structures with many knots and filaments \citep[see Figure~\ref{ngc6302} and][]{gorn00}. Such knots may be associated with the complex turbulent velocities found by \citet{acke02} in the lobes of the majority of [WC]-type PNe. However the knots observed in NGC~6302 do not appear to be H-deficient \citep{szys11}, unlike those observed in other H-poor PNe (e.g. Abell~30 and Abell~78). This could indicate that the knots in NGC~6302 were ejected during an earlier, H-rich phase of mass-loss and have since been disrupted and accelerated by a later, H-poor mass-loss event.

\subsection{Evolutionary tracks for late thermal pulse evolution}
\label{s-tracks}

\begin{figure*}
\begin{center}
\includegraphics[width=200pt, angle=270]{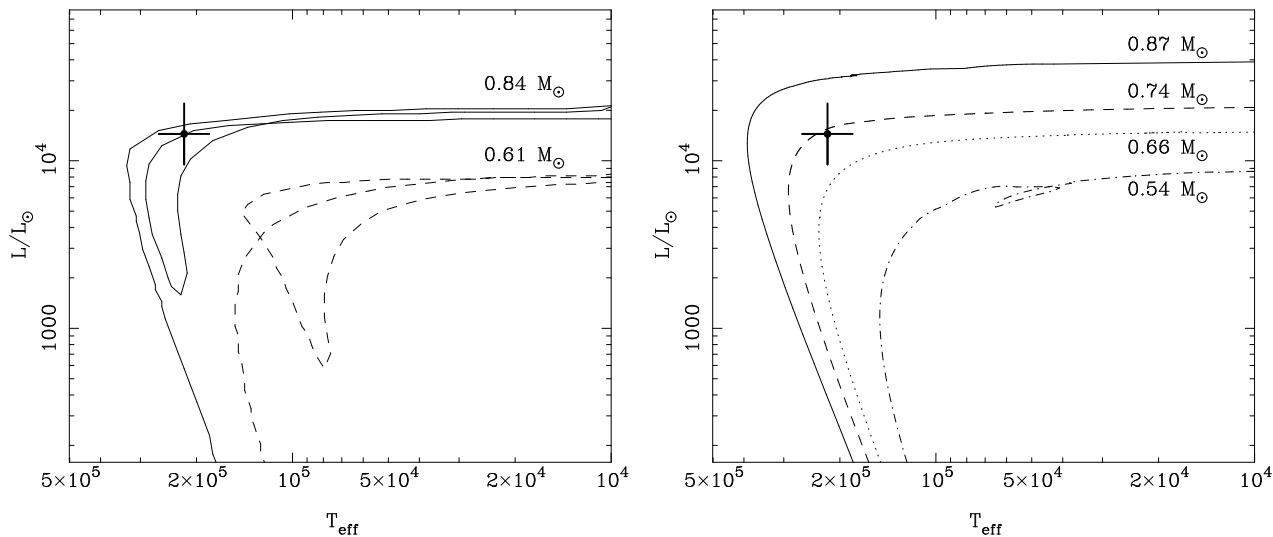}
\caption{Hertzprung-Russell diagrams for very late thermal pulse evolutionary models compared to the position of the central star of NGC~6302 derived from our models (with 20\% error bars). {\it Left:} Evolutionary tracks from \citet{bloc95b} with core masses indicated. Tracks show the first PN central star phase, the born-again event and the second PN central star phase. 
{\it Right:} Evolutionary tracks from \citet{mill06} with core masses (indicated) showing only the second PN central star phase.}
\label{evotracks}
\end{center}
\end{figure*}

To further study the central star of NGC~6302 we have compared its properties determined from our models with those from evolutionary tracks for late thermal pulse models. This might allow us to determine the core mass of the central star, and therefore potentially its original mass, as well as its evolutionary state.

Figure~\ref{evotracks} shows evolutionary tracks for the VLTP models of \citet{bloc95b} and the LTP and VLTP models of \citet{mill06}. In the models of \citet{bloc95b} the central star of NGC~6302 falls between the 0.61 and 0.84~M$_{\odot}$ core mass tracks, at $\sim$0.82~M$_{\odot}$, corresponding to an initial mass of 4.8~M$_{\odot}$. In the \citet{mill06} models the central star corresponds to a star with a core mass of $\sim$0.73~M$_{\odot}$, an initial mass of 3.7~M$_{\odot}$, and $\log \, g = 6.6$, close to our modeled value.

We can compare the timescales of these different evolutionary tracks with the known evolutionary history of NGC~6302, such as the last major mass-loss event 2200~yrs ago \citep{meab08,szys11}. The timescales for LTP events are considerably longer than those for VLTP events, which could suggest that the VLTP event is more likely given the short timescale since the last mass-loss event. Considering the VLTP model timescales, for the models of \citet{bloc95b} for a 4.8~M$_{\odot}$ initial mass star the position corresponds to an age of $\sim$2100~yrs since the star left the end of the AGB and $\sim$300 years since the VLTP event, suggesting that the eruptive event traced by \citet{meab08} was the last mass-loss event on the AGB before the central star evolved to hotter temperatures. For the models of \citet{mill06} only ages since the VLTP event are provided, which, for a 3.7~M$_{\odot}$ initial mass star correspond to $\sim$2000~yrs since the helium flash that caused this event, considerably longer than the equivalent evolutionary period in the \citet{bloc95b} models.

If the timescales of \citet{mill06} are correct this suggests that the eruptive event identified by \citet{meab08} is associated with a VLTP event. A VLTP mass-loss event however is not thought to be particularly large. \citet{vanh07} estimated that the mass lost in the VLTP eruption of V334~Sgr might be as much as 0.01~M$_{\odot}$, at least an order of magnitude smaller than would be necessary for one pair of NGC~6302's bipolar lobes. However, \citet{alth08} found evidence that PG~1159 stars may have much thinner He-rich envelopes than previously thought and this might be due to considerable mass-loss during the VLTP event, though the helium layer is only $\sim$0.02~M$_{\odot}$ thick and this puts an upper limit on what can be lost during such an event. \citet{szys11} note that the mass and momentum of the bipolar lobes is too high for them to have been ejected in a single mass-loss event and suggest that the recent mass-loss event may have accelerated gas that was previously ejected.

The large discrepancy in timescales between the models presented here for similar VLTP events implies a significant level of uncertainty in the current models that requires more accurate evolutionary tracks. Therefore at this point it is impossible to distinguish between the two sets of models as either could imply a timescale that agrees well with the evolutionary history of NGC~6302. Combining the derived core mass (0.74--0.82~M$_{\odot}$) with the nebular mass estimated from this work (4.7~M$_\odot$; Table~\ref{properties}) implies an initial stellar mass of at least 5.5~M$_{\odot}$. The initial stellar mass of 3.7--4.8~M$_\odot$ from evolutionary tracks is not far from the 5.5~M$_\odot$ that we derive.

\subsubsection{Comparison with other known PG~1159 stars}

\begin{figure}
\begin{center}
\includegraphics[width=180pt, angle=270]{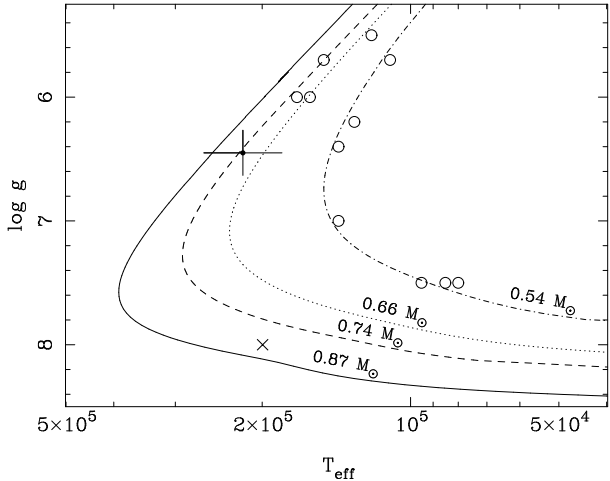}
\caption{VLTP evolutionary tracks showing the second PN central star phase in the $\log g$~-~$\log T_{eff}$ plane with core masses indicated \citep{mill06}. The central star of NGC~6302 is shown with 20\% error bars. The positions of the known PNe with PG~1159 type central stars are shown as open circles with data taken from \citet{wern06} and the helium-deficient star H1504+65 \citep{wern04} is shown with a cross.}
\label{pg1159all}
\end{center}
\end{figure}

A number of PG~1159 stars have measured effective temperatures and gravities \citep{wern06}. Figure~\ref{pg1159all} shows these stars in the $\log g$ -- $\log T_{eff}$ plane (chosen because these quantities are better constrained than their luminosities), alongside the central star of NGC~6302 (using the value of $\log g \sim 6.6$ fitted from Figure~\ref{evotracks}, which is better constrained than our modeled value), and the helium-deficient star H1504+65 \citep{wern04}. In addition we show evolutionary tracks for the \citet{mill06} LTP/VLTP models. The position of the central star of NGC~6302 is in good agreement with that of many of the other PG~1159 stars. Is is clearly hotter than any of the known PG~1159 stars, and even hotter than the previously hottest known central star, H1504+65, though that source does appear to have had a higher initial mass and a previously higher temperature (it is now on the cooling track).

\subsection{The evolutionary history of NGC~6302}

From the evidence presented here we suggest that NGC~6302 was created by a star with an initial mass of at least 5~M$_{\odot}$, in agreement with both the chemical abundances and evolutionary tracks for the central star. The star may have had a lower-mass binary companion that diverted the AGB-phase mass-loss from isotropic to equatorial, forming the circumstellar disk \citep[e.g.][]{bond90,garc04}.

The question of the age and stability of the circumstellar material is important for understanding the history of such a nebula. The presence of large dust grains \citep{hoar92} and a high fraction of crystalline material \citep{mols01} could indicate that the circumstellar disk is long-lived, providing a shielded and stable environment for these grains to build up. However, such dust characteristics could also be caused by high mass-loss rates and densities in the ejected material \citep[][found that the degree of crystallinity in OH/IR stars appears to scale with mass-loss rate]{sylv99}. \citet{pere07} measure an expansion velocity of $\sim$8~km~s$^{-1}$ for the circumstellar torus and argue that it was ejected during a $\sim$5000 year mass-loss event that started $\sim$7500 years ago. Such a scenario requires mass-loss rates of $5 \times 10^{-4}$~M$_\odot$~yr$^{-1}$ over this time period, potentially high enough to create the conditions for grain growth and crystallization in NGC~6302, but higher than current evolutionary models can sustain, even for shorter periods of time \citep[e.g.][]{vass93,bloc95b}.

The kinematic age may be a lower limit if the disk has been accelerated by radiation pressure from the highly luminous central star. A simple calculation for the momentum imparted by radiation pressure on the circumstellar disk suggests that, based on the parameters derived here and the expansion velocity measured by \citet{pere07}, this would take $\sim$500,000~years at the current stellar luminosity (assuming that $\sim$10\% of the stellar luminosity intercepts the circumstellar disk). This timescale is significantly longer than typical PNe lifetimes and as such unfeasible as the major acceleration mechanism for the circumstellar disk. A similar argument is presented by \citet{pere07} who also conclude that alternative mechanisms are required and suggest interaction with a binary companion as a likely propulsion mechanism.

The circumstellar disk would have caused later mass-loss events to form the bipolar lobes. \citet{mats05} observed the circumstellar disk to be warped, which may be caused by an interaction with a binary companion, and according to the model of \citet{icke03} would cause each pair of bipolar lobes to form with a different axis, as is observed. Hot-bottom burning at the base of the convective envelope \citep[which occurs for stars more massive than $\sim$4~M$_{\odot}$,][]{iben83,boot93}, would convert carbon to nitrogen and prevent the star or the ejecta going through a prolonged C-rich phase.

At some point the star experienced some sort of late thermal pulse, ejecting a very small amount of material, causing the central star to become H-deficient and returning to the tip of the AGB. It is not currently possible to observationally distinguish between the different LTP scenarios, though the evolutionary timescales for the VLTP models of \citet{mill06} show the strongest agreement with the known mass-loss history. Both sets of evolutionary tracks (above) suggest that the temperature of the star will continue to rise, potentially reaching a maximum temperature of $\sim$280,000~K, before entering the white dwarf cooling track again.

\section{Conclusions}

We have constructed a 3D photoionization model of the extreme planetary nebula NGC~6302 using the Monte Carlo photoionization and radiative transfer code {\sc mocassin}. Our model reproduces the majority of emission line strengths and ratios, which place strong constraints on many of the stellar and nebular properties. The model consists of a very dense circumstellar disk where densities reach 300,000~cm$^{-3}$ at the inner edge, a large pair of bipolar lobes with a constant density of 2000~cm$^{-3}$, and an intermediate component we have dubbed `the outflow'. This combination of components was required to match the majority of density- and temperature-sensitive line ratios from which a wide range of densities and temperatures have been observationally determined by previous authors. A number of line ratios remain not matched, which we suggest is due to complex density inhomogeneities throughout the nebula.

We derive a total nebular mass of 4.7~M$_\odot$, of which 1.8~M$_\odot$ is ionized, almost entirely in the bipolar lobes. The C/O ratio for NGC~6302 is 0.43 indicating a predominantly O-rich nebula, in agreement with all other measurements in the literature. Carbon is found to be marginally under-abundant compared to the solar value, while nitrogen is significantly over-abundant. The combined (C+N+O)/H abundance is 35\% larger than the solar value, all of which suggests that a 3rd dredge-up occurred in the precursor of NGC~6302, enriching the central star in carbon and nitrogen, followed by hot bottom burning CN-cycle conversion of carbon to nitrogen.

In modeling the central star we have incorporated NLTE model atmospheres to reproduce the ionizing flux. Fits to the optical emission-line spectrum imply an extremely high temperature for the central star, and we used the observed emission from high ionization-stage infrared `coronal' lines to further constrain the form of the ionizing flux distribution. Using a solar abundance stellar atmosphere we were unable to fit all the observed line fluxes, overestimating many of the lower ionization stages while underestimating the higher stages. A substantially better fit was obtained using a hydrogen-deficient stellar atmosphere, implying that the central star of NGC~6302 is likely to be H-deficient.

Finally we compare the properties of the central star with evolutionary tracks for late thermal pulses that are capable of removing the majority of hydrogen from the central star. We find a good fit to a very late thermal pulse track for a star with an initial mass of 4--5~M$_{\odot}$. Timescales for this evolutionary model imply that the central star left the top of the AGB $\sim$2100 years ago, in good agreement with the age of one of the pairs of bipolar lobes determined by \citet{meab08} to be $\sim$2200 years. This mass is in reasonable agreement with the total modeled nebular mass plus central star mass of 5.5~M$_\odot$, taking into account contributions from the dust component. The mass is also in agreement with the chemical abundances in the nebula such as the high helium abundance, low C/O ratio and slightly enhanced (C+N+O)/H ratio.

A future paper will discuss the dust component of this model that used {\sc mocassin}'s ability to fully model the radiative transfer of dust in an ionized nebula. The model includes the majority of observed amorphous and crystalline dust species, using multiple dust chemistry distributions, non-standard grain size distributions, and a new light scattering code for non-spherical dust grains.

\section{Acknowledgments}

The simulations presented in this work were performed in the DiRAC Facility jointly funded by STFC and the Large Facilities Captial Fund of BIS. The authors acknowledge support of the STFC funded Miracle Consortium (part of the DiRAC facility) in providing access to the UCL (University College London) Legion High Performance Computing Facility. The authors additionally acknowledge the support of UCL's Research Computing team with the use of the Legion facility, particularly Jeremy Yates and Dugan Witherick. This work has also made us of the CHIANTI database, a collaborative project involving researchers at NRL (USA) RAL (UK), and the Universities of: Cambridge (UK), George Mason (USA), and Florence (Italy).

We are grateful to the referee, Albert Zijlstra, for a swift and helpful report that has improved the work presented here. We would also like to thank Pete Storey, Mikako Matsuura, Jonathan Rawlings, and Tim Gledhill for interesting discussions on this work. NJW acknowledges support from the STFC and an SAO Pre-doctoral Fellowship. BE acknowledges support from an STFC Advanced Fellowship. TR us supported by the German Aerospace Center (DLR) under grant 05\,OR\,0806.

\bibliographystyle{mn2e}
\bibliography{/Users/nwright/Documents/Work/tex_papers/bibliography.bib}
\bsp

\label{lastpage}

\end{document}